\documentclass[12pt]{iopart}
\usepackage{amsfonts}
\usepackage{amssymb}
\usepackage{setstack}
\usepackage{float}
\usepackage{floatflt}
\usepackage{subfigure}
\usepackage[dvips]{graphicx}

\begin{document}
%
%

\newcommand{\Gs}{{\mathcal G}_s} 
\newcommand{\Ds}{{\mathcal D}'({\mathbb{R}^4})}  
\newcommand{\rhoo}{{\hskip 3pt {\mathop{\kern -3pt \rho}\limits^o}}} 
\newcommand{\ro}{{\hskip 3pt {\mathop{\kern -3pt r}\limits^o}}} 
\newcommand{\qd}{\hspace{1mm}}
\def\Se{S_\epsilon}
\def\Pe{P_\epsilon}
\def\bx{\mathbf x}
\newtheorem{proposition}{Proposition}

\def\Re{{\mathbb R}}
\def\css{\subset\!\subset}
\def\Nat{{\mathbb N}}
\def\E{{\mathcal E}}
\def\N{{\mathcal N}}
\def\G{{\mathcal G}}
\def\A{{\mathcal A}}
\def\L{{\mathcal L}}
\def\Sup{{\hbox{Sup}}}
\def\D{{\mathcal D}}
\def\d{{\rm d}}
\def\half{{\scriptstyle {1 \over 2}}}
\def\quarter{{\scriptstyle {1 \over 4}}}
\def\p{\partial}
\def\diag{\hbox{diag}}
\def\a{{\bf a}}
\def\b{{\bf b}}
\def\c{{\bf c}}
\def\0{{\bf 0}}
\def\1{{\bf 1}}
\def\2{{\bf 2}}
\def\3{{\bf 3}}
\def\X{{\bf X}}
\def\be{{\bf e}}

\def\CQG{Class. Quant. Grav. }
\def\GRG{J. Gen. Rel. Grav. }
\def\JMP{J. Math. Phys. }
\def\PRD{Phys. Rev. D }


\title[The thin string limit of Cosmic Strings]{The thin string limit of Cosmic Strings coupled to gravity}
\author{K R P Sj\" odin   and J A Vickers}
\address{Department of Mathematics, University of Southampton,
Southampton, SO17~1BJ, UK}
\ead{J.A.Vickers@maths.soton.ac.uk}
\begin{abstract}
The thin string limit of Cosmic Strings is investigated using a
description in terms of Colombeau's theory of 
nonlinear generalised functions. It is shown
that in this description the energy-momentum tensor has a well defined
thin string limit. Furthermore the deficit angle of the
conical spacetime that one obtains in the limit may be given in terms
of the distributional energy-momentum tensor. On the other hand it is
only in the special case of critical coupling that the energy-momentum tensor
defined in the Colombeau algebra is associated to a conventional
distribution. The asymptotics of both the matter and gravitational
field are investigated in the thin string limit and it is shown how
this leads to the `conical approximation' which is valid outside the
inner core of the string.
\end{abstract}
\submitted{\CQG}
\pacs{0420H, 0420E, 0420F, 0430N}
\maketitle
%

\section{Introduction}
In this paper we will examine the thin string limit of solutions of
the field equations for an infinite length straight Cosmic String
[described by a complex scalar field coupled to a $\mathrm{U(1)}$ gauge field]
coupled to gravity. The thin string limit has been examined by a
number of authors [1--4] 
and it is generally agreed that in the thin
string limit the metric describing an infinite length static Cosmic
String becomes that of a conical spacetime
\begin{equation}
ds^2=dt^2-dr^2-A^2r^2d\phi^2-dz^2 \label{metric1} 
\end{equation}
where $\phi$ is the usual $2\pi$-periodic coordinate of cylindrical
polar coordinates. If one changes to a new angular variable $\widetilde
\phi=A\phi$, where $\widetilde \phi \in [0,2\pi A]$, the metric becomes
that of Minkowski space so we see that (1) represents a conical metric
with angular deficit $\Delta\phi=2\pi(1-A)$. If one also makes the
weak-field approximation then one can also show (see e.g. Vilenkin
\cite{vilenkin}) 
that the angular deficit $\Delta\phi$ is related to the mass per
unit length $\mu$ of the the string in a simple way:
\begin{equation}
\Delta\phi=8\pi\mu 
\end{equation}
(where we are using units in which $\hbar=c=G=1$).

\medskip\noindent
This result is consistent with the study of ``simple line sources'' in
general relativity made by Israel \cite{israel}. He showed that if one has a
1-parameter family of spacetimes with the matter tending to a simple
line source in the limit then, providing the radial stress becomes
negligible compared to the mass density, the energy-momentum tensor
becomes that of an infinitely thin wire with distributional stress
energy. This analysis was applied by Linet \cite{linet} to a critically coupled
Cosmic String (i.e. a string for which the scalar and vector fields
have equal masses) who showed that equation (2) is satisfied in this
case. However as pointed out by Garfinkle \cite{garfinkle}, in the non-critical
case Israel's condition is violated and there fails to be a simple
relationship between $\Delta\phi$ and $\mu$ in the thin string
limit. Futamase and Garfinkle \cite{futamase}
argue that because the relationship
between $\Delta\phi$ and $\mu$ depends upon the nature of the matter
which makes up the string there is in general no well defined thin
string limit for Cosmic Strings in general relativity. This
interpretation agrees with the analysis of conical spacetimes by
Geroch and Traschen \cite{g&t}, who showed that different 1-parameter
families of spacetime with a conical spacetime as the limit could have
different limiting mass per unit length. Note however that if one puts
certain additional conditions on the 1-parameter families of
spacetime then one {\it does} recover formula (2) (see for example
Wilson \cite{wilson} and Clarke \etal \cite{cvw}).

\medskip\noindent
The difficulty of having a well defined thin string limit in general
relativity comes from the non-linear nature of the field
equations. For a linear theory such as electromagnetism a consistent
framework for describing concentrated sources is provided by
Schwartz's theory of distributions. This is for two reasons. Firstly
the equations are linear with respect to sources and fields so the
field equations are well defined for distributional sources. Secondly
for smooth sources close to a point charge (in the topology on the
space of distributions) the field produced is close to that produced
by a point charge. It is this second feature which means that in
electromagnetism the detailed internal structure of a concentrated
source is unimportant and only the distributional charge density that
it defines in the limit is significant.

\medskip\noindent
In this paper we will consider the thin string limit of Cosmic Strings
by making use of generalised functions belonging to the simplified
Colombeau algebra $\Gs$. The construction of the simplified Colombeau
algebra is outlined in section 2 but the key point to note here is
that it provides a consistent setting which allows one to multiply
distributions in a non-linear theory such as general relativity. Furthermore   
if two generalised sources are close (in the appropriate topology)
then the corresponding metrics (satisfying the appropriate boundary
conditions) are also close. Although $\Gs$ contains distributions $\Ds$
as a vector subspace, the algebra has a much richer structure because
it is able to encode non-linear information. Thus one can use these
algebras to model the thin string limit of a Cosmic String coupled to
gravity because one is able to retain certain features of the source
in the limit. This explains why one can use them to define the thin
string limit even in case when there is not a simple relationship
between $\Delta\phi$ and $\mu$.

\medskip\noindent
The plan of the paper is as follows. In section 2 we outline the
construction of the simplified Colombeau algebra. In section 3 we
illustrate the ideas by considering the thin string limit of a static
Cosmic String in Minkowski space. In section 4 we consider the thin
string limit of a static Cosmic String coupled to the gravitational
field and in section 5 examine the relationship between the angular 
deficit and mass per unit length. In section 6 we introduce the conical
approximation and in section 7 we briefly consider the case of a
dynamic Cosmic String coupled to gravity. 

\section{A brief review of Colombeau theory}
In this section we give a brief introduction to the Colombeau
algebras. For more details see \cite{col1} and \cite{biagioni}. 
The basic idea is to consider generalised functions as 1-parameter
families of smooth functions $\{f_{\epsilon}\}$. Our basic space will
thus be
\begin{equation}
\E(\Re^n)=\{\{f_{\epsilon}\}: \ 0<\epsilon<1,\ \ f_\epsilon \in
C^\infty(\Re^n) \}.
\end{equation}
We may represent a smooth function by the constant family
$f_\epsilon(x)=f(x)$, but for a function of finite differentiability
we may obtain a family of smooth functions by taking the convolution
with a suitable smoothing kernel or molifier $\Phi$
\begin{equation}
f_\epsilon(x)={1 \over {\epsilon^n}}\int_{\Re^n} f(y) \Phi((y-x)/\epsilon) \d^ny.
\end{equation}
However for the case of a smooth function we would like both
representations to be equivalent. Using a Taylor series expansion to
compare the difference between these expressions we are lead
to define two representations to be equivalent if they differ by a
`negligible function' which is defined as a 1-parameter family 
of functions which
on any compact set vanishes faster than any given positive power of 
$\epsilon$. Since we are trying to construct a {\it differential
algebra} we also require the derivatives of $f$ to  satisfy this
property and the resulting set $\N$ to be an ideal. Unfortunately $\N$
is not an ideal in $\E(\Re^n)$, but by restricting the space to
`moderate functions' $\E_M(\Re^n)$ which grow no faster than some
inverse power of $\epsilon$, one does have an ideal and we may define
the differential algebra $\Gs$ as the quotient. 

\medskip\noindent
{\bf Definition} (Moderate functions)
\begin{eqnarray*}
\E_M(\Re^n)&\!\!\!=&\!\!\!\left\{\{f_\epsilon\}:\  \forall K \css\Re^n, \forall \alpha \in
\Nat^n_0, \exists p\in\Nat, \exists \eta>0,\exists C>0 \hbox{ such
that} \right.\\
&\!\!\!&\!\!\!\qquad \left.\Sup|D^\alpha f_\epsilon(x)| \leqslant C\epsilon^{-p} \hbox{ for }
0<\epsilon <\eta \right\}.\\
\end{eqnarray*}
Note $K \css \Re^n$ indicates that $K$ is {\it compact} and we have
also employed the standard multi-index notation for $D^\alpha f$.

\medskip\noindent
{\bf Definition} (Negligible functions)
\begin{eqnarray*}
\N(\Re^n)&\!\!\!=&\!\!\!\left\{\{f_\epsilon\}:\  \forall K \css\Re^n, \forall \alpha \in
\Nat^n_0, \forall q\in\Nat, \exists \eta>0,\exists C>0 \hbox{ such
that} \right.\\
&\!\!\!&\!\!\!\left. \qquad\Sup|D^\alpha f_\epsilon(x)| \leqslant C\epsilon^{q} \hbox{ for }
0<\epsilon <\eta \right\}.\\
\end{eqnarray*}
{\bf Definition} (Simplified algebra)
\begin{displaymath}
\Gs(\Re^n)=\E_M(\Re^n)/\N(\Re^n).
\end{displaymath}
Thus a non-linear generalised function is represented by a moderate
sequence of smooth functions modulo a negligible sequence. The space
$\E_M(\Re^n)$ is a differential algebra with pointwise operations and
since the space of negligible functions is a differential ideal, $\G$
is also a commutative differential algebra. As remarked earlier one of
the advantages of the Colombeau approach is that one may frequently
interpret the results in terms of distributions using the concept of
association or weak equivalence. An element  $[f_\epsilon]$ of $\G$
is said to be associated to a distribution 
$T \in \D'$ if a representative  $\{f_\epsilon\}$ in $\E_M$ satisfies 
\begin{equation}
\forall \phi \in \D, \quad \lim_{\epsilon \to 0}\int_{\Re^n}
f_\epsilon(x)\phi(x)\d^nx=\langle T,\phi\rangle
\end{equation}
and we then write $[f_\epsilon] \approx T$.
Note that not all elements of $\G$ are associated to distributions. 

\medskip\noindent
More generally we say $[f_\epsilon] \approx [g_\epsilon]$ if
\begin{equation}
\forall \phi \in \D, \quad \lim_{\epsilon \to 0}\int_{\Re^n}
f_\epsilon(x)\phi(x)\d^nx=\lim_{\epsilon \to 0}\int_{\Re^n}
g_\epsilon(x)\phi(x)\d^nx.
\end{equation}
Association is an equivalence relation which respects addition and
differentiation. It also respects multiplication by {\it smooth} functions
but by the Schwartz impossibility results cannot respect multiplication
in general.

\medskip\noindent
The algebra presented above is the simplest of the Colombeau
algebras. Furthermore the notion of smooth function and the growth
conditions in the definitions of moderate and negligible functions
only depend upon the differential structure. This allows the simplified
algebra to be readily generalised to arbitrary manifolds. However it
does suffer from the disadvantage that the embedding of $C^p$
functions and distributions is not canonical. Thus one has to appeal
to mathematical or physical arguments outside the theory to justify a
particular representation of a non-smooth function. For the case of
the field equations for a Cosmic String coupled to gravity, we will
see in the next section that there is a natural scaling of the
coupling constants which 
can be used  as the Colombeau parameter. In other cases one can embed
by taking the convolution with an element of some special function
space and only use general properties of elements of the space in
proving results. However this is not always satisfactory and in
certain situations  the full Colombeau algebra is preferable. 
This involves enlarging the algebra substantially by making 
the generalised functions depend upon the molifier. This allows one to 
construct a canonical embedding of distributions but at the expense of 
problems with coordinate invariance. A manifestly coordinate
independent construction of the full algebra is given in \cite{advances}.
\section{The thin string limit in Minkowski space}
We consider Cosmic Strings which are described by a complex scalar
field $\Phi$ which is coupled to a vector field $A_\mu$. The equations
of motion for these fields may be derived from the Lagrangian
\begin{equation}
L = D_\mu \Phi\overline{D^\mu \Phi}-V(|\Phi|)-\frac{1}{4}F_{\mu\nu}F^{\mu\nu}
\end{equation}
where $D_\mu=\nabla_\mu=ieA_\mu$ is the gauge covariant derivative
with coupling constant $e$ and $F_{\mu\nu}=\nabla_\mu A_\nu-\nabla_\nu
A_\mu$. We write $\Phi$ in the form $\Phi=\frac{1}{\sqrt 2} S
e^{i\psi}$ and consider the case of a potential given by
$V=4\lambda(S^2-\eta^2)^2$. Then the Lagrangian becomes
\begin{equation} \label{S1}
L=\frac{1}{2}\nabla_\mu S\nabla^\mu S +
    \frac{1}{2} S^2(\nabla_\mu\psi+eA_\mu)(\nabla^\mu\psi+eA^\mu) -
    \lambda(S^2-\eta^2)^2 - \frac{1}{4}F_{\mu\nu}F^{\mu\nu}. \label{lag}
\end{equation}
Here $\lambda$ is a coupling constant and $\eta^2$ is twice the vacuum
expectation value of the scalar field. For the above Lagrangian the
mass term for the scalar field is given by $m_S^2=8\lambda\eta^2$
while that for the vector field is given by $m_P^2=e^2\eta^2$ so that
the mass terms are equal and the string is critically coupled  when
$e^2=8\lambda$. The energy-momentum tensor corresponding to the
Lagrangian (\ref{S1}) is given by
\begin{equation} \label{S5}
    T_{\mu\nu} = \nabla_\mu S\nabla_\nu S + S^2(\nabla_\mu\psi + eA_\mu)(\nabla_\nu\psi + eA_\nu) - F_\mu{}^\tau F_{\nu\tau} - g_{\mu\nu}L.
\end{equation}
We now look for solutions describing an infinite length static
cylindrically symmetric Cosmic String in Minkowski space. Following
Garfinkle \cite{garfinkle} we therefore assume that the scalar and gauge fields
have the form
\begin{eqnarray} 
\Phi=S(\rho)e^{i n\phi} \label{S6a}\\
A_{\mu}=\frac{n}{e}[P(\rho)-1]\nabla_{\mu}\phi \label{S6b}
        \end{eqnarray}
where $(t,z,\rho,\phi)$ are the usual cylindrical polar coordinates
for Minkowski space. The number $n$ is the winding number for the
string which  for simplicity we will take to be one in the remainder
of the paper. With this choice the equations of motion are given by
\begin{eqnarray}
\rho\frac{d}{d\rho}\biggl(\rho\frac{dS}{d\rho}\biggr) 
= S[4\lambda\rho^2(S^2-\eta^2)+P^2] \label{S7}\\
\rho\frac{d}{d\rho}\biggl(\rho^{-1}\frac{dP}{d\rho}\biggr) 
= e^2S^2P. \label{S8}
\end{eqnarray}
The requirement that the solution has finite energy implies that
\begin{eqnarray} 
S(\rho) \to \eta, \quad \hbox{as} \quad \rho \to \infty \label{B1} \\
P(\rho) \to 0, \quad \hbox{as} \quad \rho \to \infty. \label{B2} 
\end{eqnarray}
On the other hand at the centre the solution must smoothly attain the
symmetric state which requires
\begin{eqnarray} 
S(0)=0 \label{B3} \\
P(0)=1. \label{B4}
\end{eqnarray}
The fields describing the Cosmic String are therefore solutions of (\ref{S7}),
(\ref{S8}) satisfying (\ref{B1})--(\ref{B4}).

\medskip\noindent
With the fields taking the form (\ref{S6a}), (\ref{S6b}) 
the energy-momentum tensor is
diagonal and is given in terms of $S$ and $P$ by
    \begin{equation}
    T_{\mu\nu} = \sigma\nabla_\mu t\nabla_\nu t + 
p_z\nabla_\mu z\nabla_\nu z + 
p_\rho\nabla_\mu \rho\nabla_\nu \rho + \rho^2p_\phi\nabla_\mu 
\phi\nabla_\nu \phi
    \end{equation}
where
    \begin{eqnarray}
    \sigma = -p_z = \half\biggl[\biggl(\frac{dS}{d\rho}\biggr)^2
+ \frac{S^2P^2}{\rho^2} + 2\lambda(S^2-\eta^2)^2 +
\frac{1}{e^2\rho^2}\biggl(\frac{dP}{d\rho}\biggr)^2\biggr]
\label{sigma} \\
    p_\rho = \half\biggl[\biggl(\frac{dS}{d\rho}\biggr)^2 
- \frac{S^2P^2}{\rho^2} - 2\lambda(S^2-\eta^2)^2 + 
\frac{1}{e^2\rho^2}\biggl(\frac{dP}{d\rho}\biggr)^2\biggr] \label{prho}\\
    p_\phi = \half\biggl[-\biggl(\frac{dS}{d\rho}\biggr)^2 + 
\frac{S^2P^2}{\rho^2} - 2\lambda(S^2-\eta^2)^2 
+ \frac{1}{e^2\rho^2}\biggl(\frac{dP}{d\rho}\biggr)^2\biggr]. \label{pphi}
    \end{eqnarray}
The mass per unit length in Minkowski space is therefore given by
\begin{equation}
\mu=\int_{\Re^2}\sigma \d^2x=2\pi\int_{0}^\infty \sigma\rho \d\rho
\label{mass}
\end{equation}
where $\sigma$ is given by (\ref{sigma}).

\medskip\noindent
Although the fields extend to infinity, they decay exponentially
fast. For large $\rho$ the asymptotic behaviour is given by
\begin{eqnarray}
P(\rho)\sim \rho K_1(e\eta\rho)\sim\sqrt{\frac{\pi}{2e\eta}}\rho^{1/2}e^{-e\eta\rho}\\
S(\rho)\sim \eta-K_0(\sqrt{8\lambda}\eta\rho)\sim\eta-\sqrt{\frac{\pi}{\sqrt{32\lambda}\eta}}\rho^{-1/2}e^{-\sqrt{8\lambda}\eta\rho}.
\end{eqnarray}
so that $\sigma$ also decays exponentially fast. We therefore define
an ``effective radius'' $\rhoo$ for the Cosmic String by requiring
that 90\% of the total mass per unit length is within this radius and hence
\begin{equation}
\int_0^\rhoo\sigma(\rho)\rho \d\rho=
\frac{9}{10}\int_0^\infty\sigma(\rho)\rho \d\rho. \label{effective}
\end{equation}
Note that the effective
radius is roughly the Compton wavelength of the scalar field which for
typical values of the coupling constants is  
of the order of 300 Planck lengths. If we define
$\ro=\sqrt\lambda\eta\rhoo$, then numerical investigations show that
$\ro$ is very close to 1 as $e$, $\lambda$ and $\eta$ vary over
several orders of magnitude. 

\medskip\noindent
Thus in practice most of the matter is highly concentrated on the axis
of the string. For this reason it is natural to try and model a Cosmic
String by taking a thin string limit in which all the matter is
concentrated on the axis. However as noted by a number of authors (see eg
\cite{cvw}) 
there are problems in simply using a $\delta$-function to model
the source. The approach adopted in this paper is to allow $e$
$\lambda$ and $\eta$ to depend upon a parameter $\epsilon$ so that
\begin{displaymath}
e=e(\epsilon), \qquad \lambda=\lambda(\epsilon), \qquad
\eta=\eta(\epsilon), 
\end{displaymath}
with the physical values given by $\epsilon=1$ and the thin string
limit given by letting $\epsilon$ tend to zero. We then consider the
corresponding 1-parameter family of solutions $\Se(\rho)$ and
$\Pe(\rho)$ of (\ref{S7})--(\ref{B4}) which we regard as elements of 
$\E_M$, whose equivalence classes $[\Se]$ and $[\Pe]$ are elements of
$\Gs$. Since $\Gs$ is a differential algebra we may compute the
energy-momentum tensor (and other physically interesting quantities)
within $\Gs$. Whether these quantities have an interpretation in terms
of the usual theory of distributions will depend upon whether they are
associated to a conventional distribution. 

\medskip\noindent
We now consider the properties we require of the thin string limit in
order that it retains the important features of the the physics. In
order that $\epsilon \to 0$ represents the thin string limit we
require that the effective radius $\rhoo(\epsilon)$ defined by
(\ref{effective}) scales linearly with $\epsilon$, while the mass per
unit length $\mu(\epsilon)$ is independent of $\epsilon$. We also
require that the relative strength of the coupling between the scalar
and vector field given by $e^2(\epsilon)$ compared to the self-coupling
of the scalar field given by $\lambda(\epsilon)$ is independent of
$\epsilon$. We therefore try and choose the scaling for $e(\epsilon)$,
$\lambda(\epsilon)$ and $\eta(\epsilon)$ so that
\begin{eqnarray}
\rhoo(\epsilon)=\epsilon\rhoo(1)=\epsilon\rhoo \label{i}\\
\mu(\epsilon)=\mu(1)=\mu \label{ii} \\
e^2(\epsilon)/\lambda(\epsilon)=e^2(1)/\lambda(1)=\alpha. \label{iii}
\end{eqnarray}
In fact conditions (\ref{i})--(\ref{iii}) uniquely determine the
$\epsilon$ dependence of $e(\epsilon)$, $\lambda(\epsilon)$ and 
$\eta(\epsilon)$ to be given by
\begin{eqnarray}
e(\epsilon)=e/\epsilon \label{ib} \\
\lambda(\epsilon)=\lambda/\epsilon^2 \label{iib} \\
\eta(\epsilon)=\eta. \label{iiib}
\end{eqnarray}
To see this we start by observing that after a suitable rescaling the
only significant parameter in (\ref{S7}) and (\ref{S8}) is the ratio
of the coupling given by $\alpha$. Thus a solution to (\ref{S7}) and
(\ref{S8}) satisfying (\ref{B1})--(\ref{B4}) is given by
\begin{eqnarray}
S(\rho)=\eta(\epsilon) X({\sqrt{\lambda(\epsilon)}}\eta(\epsilon)\rho) 
\label{a} \\
P(\rho)=Y({\sqrt{\lambda(\epsilon)}}\eta(\epsilon)\rho) \label{b} 
\end{eqnarray}
where $X(r)$ and $Y(r)$ are solutions of 
    \begin{eqnarray}
    r\frac{d}{dr}\biggl(r\frac{dX}{dr}\biggr) =
    X[4\lambda r^2(X^2-1)+Y^2] \label{7}\\
    r\frac{d}{dr}\biggl(r^{-1}\frac{dY}{dr}\biggr) = \alpha X^2Y \label{8}
    \end{eqnarray}
satisfying the boundary conditions
\begin{eqnarray} 
X(0)=0, \qquad \lim_{r \to \infty}X(r)=1 \label{9}\\
Y(0)=1, \qquad \lim_{r \to \infty}Y(r)=0. \label{12}
\end{eqnarray}
Substituting for $S(\rho)$ and $P(\rho)$ using (\ref{a}) and (\ref{b})
in the expression (\ref{sigma}) for $\sigma$ we find that 
\begin{equation}
\sigma_\epsilon(\rho)=\lambda(\epsilon)\eta^4(\epsilon) 
h({\sqrt{\lambda(\epsilon)}}\eta(\epsilon)\rho) \label{sigma2a}
\end{equation}
where
\begin{equation}
h(r)=\half\biggl\{[X'(r)]^2+\frac{1}{r^2}X^2(r)Y^2(r)+2[X^2(r)-1]^2+\frac{1}{\alpha
r^2}[Y'(r)]^2\biggr\}.
\end{equation}
The corresponding mass per unit length is then given by
\begin{eqnarray}
\mu_\epsilon &= 2\pi\int_{0}^\infty \sigma_\epsilon(\rho)\rho \d\rho \nonumber\\
    &= 2\pi\int_{0}^\infty \eta^2(\epsilon) h(r) r \d r \nonumber\\
    &=\eta^2(\epsilon) M
\end{eqnarray}
where 
\begin{equation}
M=2\pi\int_0^\infty h(r)r \d r.
\end{equation}
Since $M$ is independent of $\epsilon$ we see that for the mass per
unit length to be independent of $\epsilon$ we also require that
$\eta(\epsilon)$ is independent of $\epsilon$ so that
\begin{equation}
\eta(\epsilon)=\eta(1)=\eta.
\end{equation}
We next introduce a scaling factor $s(\epsilon)$ defined by
\begin{equation}
s(\epsilon)=\frac{e(\epsilon)}{e(1)}=\frac{e(\epsilon)}{e}.
\end{equation}
Then in order to make $e^2(\epsilon)/\lambda(\epsilon)$ independent of
$\epsilon$ we see that $\lambda(\epsilon)$ must scale according to
\begin{equation}
\lambda(\epsilon)=s^2(\epsilon)\lambda.
\end{equation}
We now calculate how the effective radius depends upon
$s(\epsilon)$. We first let $\mu(\rho)$ be the total mass per unit
length within a radius $\rho$ so that
\begin{eqnarray}
\mu_\epsilon(\rho)&=2\pi\int_{0}^\rho \sigma_\epsilon(x)x \d x \nonumber\\
&=2\pi\eta^2\int_{0}^{s(\epsilon){\sqrt\lambda}\eta\rho} h(r)r \d r \nonumber\\
&=H(s(\epsilon)\rho)
\end{eqnarray}
where $H(r)$ is given by
\begin{equation}
H(r)=2\pi\eta^2\int_0^{{\sqrt\lambda}\eta r} h(s) s \d s.
\end{equation}
Now when $\rho=\rhoo_\epsilon$ we require
\begin{displaymath}
\mu_\epsilon(\rhoo_\epsilon)=\frac{9}{10}\mu_\epsilon=\frac{9}{10}\mu
\end{displaymath}
so that we want
\begin{displaymath}
H(s(\epsilon)\rhoo_\epsilon)=\frac{9}{10}\mu=\mathrm{const}.
\end{displaymath}
But by requirement (\ref{i}) $\rhoo_\epsilon=\epsilon\rhoo$ so that we
need
\begin{equation}
H(s(\epsilon)\epsilon\rhoo)=\mathrm{const}.
\end{equation}
which implies that $s(\epsilon)\epsilon$ is constant (since H is
increasing) and since we require $s(1)=1$ we must have
\begin{equation}
s(\epsilon)=\frac{1}{\epsilon}.
\end{equation}
We have therefore established the following proposition.

\begin{proposition}
The $\epsilon$-dependence of $e(\epsilon)$, $\lambda(\epsilon)$ and
$\eta(\epsilon)$ such that a solution of the field equations for a
Cosmic String in Minkowski space satisfies 
conditions (\ref{i})--(\ref{iii}) is uniquely given by
\begin{displaymath}
e(\epsilon)=e/\epsilon, \qquad \lambda(\epsilon)=\lambda/\epsilon^2, \qquad 
\eta(\epsilon)=\eta
\end{displaymath}
and the corresponding solutions of (\ref{S7}) and (\ref{S8})
satisfying (\ref{B1})--(\ref{B4}) are given by
\begin{eqnarray}
&S_\epsilon(\rho)=\eta X({\sqrt\lambda}\eta\rho/\epsilon) \label{c} \\
&P_\epsilon(\rho)=Y({\sqrt\lambda}\eta\rho/\epsilon) \label{d}
\end{eqnarray}
where $X(r)$ and $Y(r)$ are solutions of (\ref{7}) and (\ref{8})
satisfying the boundary conditions (\ref{9})--(\ref{12}).
\end{proposition}
We next calculate the corresponding value of the energy density
$\sigma_\epsilon(\rho)$. Substituting (\ref{c}) and (\ref{d}) into
the expression for $\sigma$ (\ref{sigma}) we find
\begin{equation}
\sigma_\epsilon(\rho)=\frac{\lambda\eta^4}{\epsilon^2}
h\biggl(\frac{{\sqrt\lambda}\eta\rho}{\epsilon}\biggr) =
\frac{1}{\epsilon^2}\sigma(\rho/\epsilon).
\end{equation}
We now show that $\sigma_\epsilon(\rho)$ represents a
$\delta$-function within $\Gs$. Because of the coordinate singularity
at the origin in polar coordinates we first transform to Cartesian
coordinates and define
$\widetilde\sigma_\epsilon(x,y)=\sigma_\epsilon(\sqrt{x^2+y^2})$ and
$d_\epsilon(x,y)=(1/\mu)\widetilde\sigma_\epsilon(x,y)$. Then
\begin{eqnarray}
\int_{\Re^2}d_\epsilon(x,y)\d x\d y&=\frac{1}{\mu}\int_{\Re^2}\tilde\sigma_\epsilon(x,y)\d x\d y
\nonumber \\
&=\frac{2\pi}{\mu}\int_0^\infty \sigma_\epsilon(\rho)\rho \d\rho \nonumber\\
&=\frac{\mu_\epsilon}{\mu} \nonumber\\
&=1 \quad \hbox{(since $\mu_\epsilon=\mu\quad \forall\epsilon$)}.
\end{eqnarray}
Also
\begin{eqnarray}
d_\epsilon(x,y)&=\frac{1}{\mu}\sigma_\epsilon(\sqrt{x^2+y^2}) \nonumber\\
&=\frac{1}{\epsilon^2\mu}\sigma(\sqrt{x^2+y^2}/\epsilon) \nonumber \\
&=\frac{1}{\epsilon^2}d_1(x/\epsilon,y/\epsilon).
\end{eqnarray}
Thus $d_\epsilon(\bx)$ is a 1-parameter family of smooth functions
satisfying
\begin{enumerate}
\item $\displaystyle{ \int d_\epsilon(\bx)\d\bx=1}$
\item $\displaystyle{ d_\epsilon(\bx)=\frac{1}{\epsilon^2}
d\left(\frac{\bx}{\epsilon}\right)}$.
\end{enumerate}
Using the results of section 2 we see that $[\d_\epsilon(\bx)]$
represents a $\delta$-function in $\Gs$. Note that although the
concept of delta-function in $\Gs$ is not unique, the various
delta-functions are all associated to the usual distributional
$\delta$-function so that
\begin{equation}
[\tilde\sigma_\epsilon(x,y)] \approx \mu\delta^{(2)}(x,y).
\end{equation}
Using a similar argument one can formally show that the principal pressures
$p_\rho$ and $p_\phi$ are proportional to $\delta$-functions
\begin{eqnarray}
&[p_\rho^\epsilon(x,y)] \approx a\delta^{(2)}(x,y) \label{p_r} \\
&[p_\phi^\epsilon(x,y)] \approx b\delta^{(2)}(x,y) \label{p_p}
\end{eqnarray}
where
\begin{eqnarray}
a=2\pi\int_0^\infty p_\rho(\rho)\rho \d\rho \\
b=2\pi\int_0^\infty p_\phi(\rho)\rho \d\rho. 
\end{eqnarray}
However because of the degeneracy of the polar coordinates at the
origin the meaning of (\ref{p_r}) and (\ref{p_p}) is unclear. We
therefore transform to Minkowski coordinates $(t,x,y,z)$ and find that
the corresponding distributional energy-momentum tensor $\widehat T_{\mu\nu}$ is 
given in these coordinates by
\begin{equation}\fl
\widehat T_{\mu\nu}=\left(
\begin{array}{cccc}
\mu & 0 & 0 & 0 \\
 0  & c & 0 & 0 \\
 0  & 0 & c & 0 \\
 0  & 0 & 0 & \mu
\end{array}\right)\delta^{(2)}(x,y)+d\left(   
\begin{array}{cccc}
0 & 0 & 0 & 0 \\
 0  & \frac{x^2-y^2}{x^2+y^2} & \frac{2xy}{x^2+y^2} & 0 \\
 0  & \frac{2xy}{x^2+y^2} & \frac{y^2-x^2}{x^2+y^2} & 0 \\
 0  & 0 & 0 & 0
\end{array}\right)\delta^{(2)}(x,y)
\label{cart}
\end{equation}
where $c=\frac{1}{2}(a+b)$ and $d=\frac{1}{2}(c+d)$.
We note that $\frac{x^2-y^2}{x^2+y^2}=\cos2\theta$ and 
$\frac{2xy}{x^2+y^2}=\sin2\theta$ are bounded but have a directional
dependent limit. Indeed the functions $\cos2\theta$ and $\sin2\theta$
are precisely the directional dependent functions that appear in the
conical metric when written in Cartesian coordinates (see for example
\cite{cvw}). More importantly in the classical theory of distributions
one can only multiply elements of $\Ds$ by {\it smooth} functions so
that the entries in the second matrix in  (\ref{cart}) do not define
elements of $\Ds$. It is not hard to extend the notion of
multiplication in $\Ds$ in some ad hoc way to allow these sort of
products (see for example Oberguggenberger \cite{mo}) but instead we
prefer to calculate the energy-momentum tensor within $\Gs$. To do
this we {\it first} calculate the ($\epsilon$-dependent)
energy-momentum tensor in Minkowski coordinates for the 1-parameter
family of solutions given by (\ref{c}) and (\ref{d}). This tensor is
regular (even on the axis) and the components define elements of
$\E_M$ whose equivalence classes define elements of $\Gs$. Thus
$[T^\epsilon_{\mu\nu}(t,x,y,z)]$ for $\mu,\nu =0,\ldots 3$ is a well
defined element of $\Gs$. One can then show, for example that
\begin{equation}
\big[T^\epsilon_{xy}\big] \approx 
\big[\frac{2dxy}{x^2+y^2}\delta^{(2)}(x,y)\big]  
\end{equation}
where the product on the right hand side is within $\Gs$ (and hence
well defined). 

\subsection{Critical coupling}
An important special case of the field equations occurs for critical
coupling, where the masses of the scalar and vector fields are
equal. With our choice of numerical factors in the definition of the
fields and coupling constants, critical coupling occurs when
$e^2=8\lambda$ (or $\alpha=8$). In this case one can show that the
field equations reduce to the Bogomol'nyi equations \cite{shellard}
which in cylindrical symmetry are given by
\begin{eqnarray}
&\frac{dX}{dr}=\frac{XY}{r} \label{e} \\
&\frac{dY}{dr}=-4r(1-X^2). \label{f}
\end{eqnarray}
Substitution into (\ref{prho}) and (\ref{pphi}) shows that for the
case of critical coupling 
\begin{eqnarray}
&p_\rho=0 \\
&p_\phi=0
\end{eqnarray}
and thus $a$, $b$, and hence $c$ and $d$ vanish. Furthermore in the
case of critical coupling one may integrate (\ref{sigma}) by parts and
use (\ref{e}) and (\ref{f}) to show that for winding number $n$ the mass per unit length is given by
\begin{equation}
\mu=\pi n\eta^2
\end{equation}
which leads to 
\begin{equation}
[T^\epsilon_{\mu\nu}(t,x,y,z)] \approx \mu\delta^{(2)}(x,y)\hbox{diag}(1,0,0,1).
\end{equation}
Thus in the case of critical coupling the distributional
energy-momentum tensor takes the form that is usually used in the thin
string limit. As we have seen above this is a valid choice for a
critically coupled Cosmic String where $p_\rho$ and $p_\phi$
vanish and for such strings it is reasonable to describe the
energy-momentum tensor using conventional $\delta$-functions which do
not model any internal structure. However for {\it non-critical}
coupling the energy-momentum tensor is more complicated and cannot be
modeled adequately using conventional distributions but can be
described using non-linear functions belonging to the simplified
Colombeau algebra. Because of the `micro-local' structure of elements
of this algebra one can take the thin string limit but still retain
some features of the internal structure  of the source in the limiting
energy-momentum tensor.

\section{Static Cosmic Strings coupled to gravity}
In this section we examine the physically more important case of a
Cosmic String coupled to gravity. We again consider the case of an
infinite length static Cosmic String, so the Lagrangian is still given
by (\ref{S1}), with $\Phi$ and $A_\mu$ taking the form given by 
(\ref{S6a}) and (\ref{S6b}), but with the indices in the Lagrangian
raised using a general static cylindrically symmetric metric rather
than that of Minkowski space used in the previous section. Such a
metric may be written in the form
\begin{equation}
ds^2=e^{a(\rho)}(d\tau^2-d\rho^2)-\rho^2e^{b(\rho)}d\phi^2
-e^{c(\rho)}dz^2.
\label{metric}
\end{equation}
This is a special case of the generalised Jordan, Ehlers, Kundt and
Kompaneets metric (with the twist $\omega=0$) used by Sj\" odin \etal
\cite{SSV1} in analysing dynamic Cosmic Strings. The two metrics are
related by taking $a=2(\gamma-\psi)$, $b=-2\psi$ and
$c=2(\psi+\mu)$. Note also that $a$, $b$ and $c$ are slightly different from
the form chosen by Garfinkle \cite{garfinkle} in his analysis.

\medskip\noindent
We now define four orthonormal covector fields according to
\begin{equation}
\widehat \tau_\mu=e^{a/2}\delta^0_\mu, \quad 
\widehat \rho_\mu=e^{a/2}\delta^1_\mu, \quad 
\widehat \phi_\mu=\rho e^{b/2}\delta^2_\mu, \quad 
\widehat z_\mu=e^{c/2}\delta^3_\mu.
\end{equation}
The energy-momentum tensor is given by (\ref{S5}) which with the choice
(\ref{S6a}), (\ref{S6b}) and the form of metric (\ref{metric}) is given by
\begin{equation}
    T_{\mu\nu} = \sigma \widehat{\tau}_\mu \widehat{\tau}_\nu 
+ p_\rho \widehat{\rho}_\mu \widehat{\rho}_\nu 
        + p_\phi \widehat{\phi}_\mu \widehat{\phi}_\nu + p_z \widehat{z}_\mu \widehat{z}_\nu
\end{equation}
where
\begin{eqnarray}
\fl
  &\sigma = -p_z =
  \half\biggl[e^{-a}\biggl(\frac{dS}{d\rho}\biggr)^2 
+\frac{S^2P^2}{\rho^2 e^b} + 2\lambda(S^2-\eta^2)^2 +
  \frac{e^{-a}}{e^2\rho^2 e^b}\biggl(\frac{dP}{d\rho}\biggr)^2\biggr] 
\label{C1}\\
&p_\rho = \half\biggl[e^{-a}\biggl(\frac{dS}{d\rho}\biggr)^2 - 
\frac{S^2P^2}{\rho^2 e^b} - 2\lambda(S^2-\eta^2)^2 
+ \frac{e^{-a}}{e^2\rho^2 e^b}\biggl(\frac{dP}{d\rho}\biggr)^2\biggr]
\label{C2} \\
&p_\phi = \half\biggl[-e^{-a}\biggl(\frac{dS}{d\rho}\biggr)^2 
+ \frac{S^2P^2}{\rho^2 e^b} - 2\lambda(S^2-\eta^2)^2 
+ \frac{e^{-a}}{e^2\rho^2
e^b}\biggl(\frac{dP}{d\rho}\biggr)^2\biggr].
\label{C3}
\end{eqnarray}

\medskip\noindent
Before considering the full field equations we consider a combination
of Einstein equations which allows us to simplify the metric. We find
\begin{equation}
R_{\mu\nu}(\widehat \tau^\mu\widehat \tau^\nu+\widehat z^\mu\widehat z^\nu)
=8\pi(T_{\mu\nu}-\half g_{\mu\nu})
(\widehat \tau^\mu\widehat \tau^\nu+\widehat z^\mu\widehat z^\nu)=0
\end{equation}
since $\sigma=-p_z$. If we calculate the left hand side in terms of
$a$, $b$ and $c$ we find
\begin{equation}
    \frac{d^2}{d\rho^2}(a-c) +
\frac{d}{d\rho}(a-c)\biggl[\frac{1}{\rho}
+\frac{1}{2}\cdot\frac{d}{d\rho}(b+c)\biggr] = 0.
\end{equation}
which implies that
\begin{equation}
    \frac{d}{d\rho}(a-c) = E \frac{e^{-\frac{1}{2}(b+c)}}{\rho}
\label{7.12}
\end{equation}
where $E$ is a constant. Regularity on the $z$-axis implies that $E=0$
so that the relation (\ref{7.12}) implies that $a=c+\hbox{const.}$ 
Without loss of
generality we may rescale $\rho$ and $z$ so that $a(0)=c(0)=0$ and
hence 
\begin{equation}
a=c \qquad \hbox{$\forall \rho$.}
\end{equation}
Note that $c$ does not appear in equations
(\ref{C1})--(\ref{C3}), so the expression for the energy-momentum tensor
is unchanged by this simplification to the metric.

\medskip\noindent
The boundary conditions for the matter fields  are the same as those
in Minkowski space and are given by (\ref{B1})--(\ref{B4}). For the
metric, elementary flatness on the axis together with the choice
$a(0)=0$ requires that $b(0)=0$, while the stronger condition of
$C^2$-regularity implies that $a'(0)=0$ and $b'(0)=0$
\cite{clarke&wilson}.

\subsection{The thin string limit}

\medskip\noindent
As in the case of a Cosmic String in Minkowski space one may rescale
the equations to obtain differential equations involving just the ratio
$\alpha=e^2/\lambda$ rather than $e$ and $\lambda$ separately. 
We therefore introduce the rescaled field and metric variables $X$,
$Y$, $A$ and $B$, which are defined by
\begin{eqnarray}
S(\rho)=\eta X(\sqrt\lambda \eta\rho), \\
P(\rho)= Y(\sqrt\lambda \eta\rho), \\
a(\rho)=c(\rho)=A(\sqrt\lambda \eta\rho), \\
b(\rho)=B(\sqrt\lambda \eta\rho).
\end{eqnarray}
We also follow Garfinkle \cite{garfinkle} and further simplify the equations by
replacing $B(r)$ by the new variable $K(r)$ where 
\begin{equation}
K(r)=re^{\frac{1}{2}(A+B)}
\end{equation}
Then $X(r)$, $Y(r)$, $A(r)$ and $K(r)$ are solutions of
\begin{eqnarray}
    &K\frac{d}{dr} \biggl(K \frac{dA}{dr}\biggr) = 
- 8\pi\eta^2e^A \biggl[2K^2(X^2-1)^2 
- \frac{1}{\alpha} \biggl(\frac{dY}{dr}\biggr)^2\biggr] \label{S35}\\[5pt]
    &K\frac{d^2K}{dr^2} = 
- 8\pi\eta^2e^A [2K^2(X^2-1)^2 + e^A X^2 Y^2] \label{S36}\\[5pt]
    &K\frac{d}{dr}\biggl(K\frac{dX}{dr}\biggr) = 
e^A X[4K^2(X^2-1) + e^A Y^2] \label{S37}\\[5pt]
    &\frac{K}{e^A}\cdot\frac{d}{dr}\biggl(\frac{e^{A}}{K}\cdot\frac{dY}{dr}
\biggr) = \alpha e^A X^2Y \label{S38}
\end{eqnarray}
satisfying the boundary conditions
\begin{eqnarray}
&X(0)=0,\qquad \lim_{r \rightarrow\infty}X(r)=1, \label{C4} \\
&Y(0)=1,\qquad \lim_{r \rightarrow\infty}Y(r)=0, \label{C5} \\
&A(0)=0,\qquad A'(0)=0, \label{C6} \\
&K(0)=0,\qquad K'(0)=1. \label{C7}
\end{eqnarray}

\medskip\noindent
One may now consider the thin string limit by allowing $e$, $\lambda$
and $\eta$ to depend upon $\epsilon$ as in the previous section. For a
string in a curved spacetime we define the effective radius $\rhoo$ to
be the {\em proper} radial distance which contains 90\% of the mass of
the string. Then using similar arguments to the flat space case one
can establish the following proposition
\begin{proposition}
The $\epsilon$-dependence of $e(\epsilon)$, $\lambda(\epsilon)$ and
$\eta(\epsilon)$ such that a solution of the field equations for a
Cosmic String coupled to gravity  satisfies 
conditions (\ref{i})--(\ref{iii}) is uniquely given by
\begin{displaymath}
e(\epsilon)=e/\epsilon, \qquad \lambda(\epsilon)=\lambda/\epsilon^2, \qquad 
\eta(\epsilon)=\eta
\end{displaymath}
and the corresponding solutions of the field equations 
satisfying (\ref{B1})--(\ref{B4}) are given by
\begin{eqnarray}
&S_\epsilon(\rho)=\eta X({\sqrt\lambda}\eta\rho/\epsilon) \label{c1} \\
&P_\epsilon(\rho)=Y({\sqrt\lambda}\eta\rho/\epsilon) \label{c2} \\
&a_\epsilon(\rho)=A({\sqrt\lambda}\eta\rho/\epsilon) \label{c3} \\
&b_\epsilon(\rho)=B({\sqrt\lambda}\eta\rho/\epsilon) \label{c4} \\
&c_\epsilon(\rho)=C({\sqrt\lambda}\eta\rho/\epsilon) \label{c5} \\
\end{eqnarray}
where $r^2e^{B(r)}=K^2(r)e^{-A(r)}$ and $X(r)$, $Y(r)$, $A(r)$ and
$K(r)$  are solutions of (\ref{S35})--(\ref{S38})
satisfying the boundary conditions (\ref{C4})--(\ref{C7}).
\end{proposition}
The corresponding energy-momentum tensor is given in the
$(\tau,\rho,\phi,z)$ coordinates by
\begin{equation}
T_{\mu\nu}^\epsilon = \sigma_\epsilon(\rho) 
\widehat{\tau}^\epsilon_\mu\widehat{\tau}^\epsilon_\nu + 
p_\rho^\epsilon(\rho) \widehat{\rho}^\epsilon_\mu \widehat{\rho}^\epsilon_\nu + 
p_\phi^\epsilon(\rho) \widehat{\phi}^\epsilon_\mu \widehat{\phi}^\epsilon_\nu + 
p_z^\epsilon(\rho) \widehat{z}^\epsilon_\mu \widehat{z}^\epsilon_\nu 
\end{equation}
where
\begin{equation}\fl
\widehat \tau^\epsilon_{\mu}=e^{A(\sqrt\lambda\eta\rho/\epsilon)}\delta^0_\mu,\quad
\widehat \rho^\epsilon_{\mu}=e^{A(\sqrt\lambda\eta\rho/\epsilon)}\delta^1_\mu,\quad
\widehat \phi^\epsilon_{\mu}=\rho
e^{B(\sqrt\lambda\eta\rho/\epsilon)}\delta^2_\mu,\quad
\widehat z^\epsilon_{\mu}=e^{C(\sqrt\lambda\eta\rho/\epsilon)}\delta^3_\mu,\quad
\end{equation}
and
\begin{eqnarray}
&\sigma_\epsilon(\rho)=\frac{\lambda\eta^2}{\epsilon^2}
\widetilde\sigma\biggl(\frac{\sqrt\lambda\eta\rho}{\epsilon}\biggr), \qquad
p_\rho^\epsilon(\rho)=\frac{\lambda\eta^2}{\epsilon^2}
\widetilde{p}_\rho\biggl(\frac{\sqrt\lambda\eta\rho}{\epsilon}\biggr), \\
&p_\phi^\epsilon(\rho)=\frac{\lambda\eta^2}{\epsilon^2}
\widetilde{p}_\phi\biggl(\frac{\sqrt\lambda\eta\rho}{\epsilon}\biggr), \qquad
p_z^\epsilon(\rho)=\frac{\lambda\eta^2}{\epsilon^2}
\widetilde{p}_z\biggl(\frac{\sqrt\lambda\eta\rho}{\epsilon}\biggr).
\end{eqnarray}
with
\begin{eqnarray}
\fl\hspace{20mm}    \widetilde\sigma = -\widetilde p_z = 
\half\eta^2\biggl[e^{-A}\biggl(\frac{dX}{dr}\biggr)^2 
+ \frac{e^AX^2Y^2}{K^2} + 2(X^2-1)^2 
+\frac{1}{\alpha K^2}\biggl(\frac{dY}{dr}\biggr)^2\biggr]\label{sigma2}\\
\fl\hspace{20mm}    \widetilde p_r = 
\half\eta^2\biggl[e^{-A}\biggl(\frac{dX}{dr}\biggr)^2 - 
\frac{e^AX^2Y^2}{K^2} - 2(X^2-1)^2 + 
\frac{1}{\alpha K^2}\biggl(\frac{dY}{dr}\biggr)^2\biggr] \label{pr}\\
\fl\hspace{20mm}    \widetilde p_\phi = 
\half\eta^2\biggl[-e^{-A}\biggl(\frac{dX}{dr}\biggr)^2 +
\frac{e^AX^2Y^2}{K^2} - 2(X^2-1)^2 + 
\frac{1}{\alpha K^2}\biggl(\frac{dY}{dr}\biggr)^2\biggr].\label{pphi2}
\end{eqnarray}
Note that if we transform to rescaled coordinates $(t,r,\phi,z)$ with
$t=\sqrt\lambda \eta\tau$ and $r=\sqrt\lambda\eta\rho$ [where
$\lambda=\lambda(1)$ and $\eta=\eta(1)$ are the {\it physical} values]
then the expression for the energy-momentum tensor becomes
\begin{equation}
T_{\mu\nu}^\epsilon = \widetilde\sigma_\epsilon(r) 
\widetilde{t}^\epsilon_\mu\widetilde{t}^\epsilon_\nu + 
\widetilde{p}_r^\epsilon(r) \widetilde{r}^\epsilon_\mu 
\widetilde{r}^\epsilon_\nu + 
\widetilde{p}_\phi^\epsilon(r) \widetilde{\phi}^\epsilon_\mu 
\widetilde{\phi}^\epsilon_\nu + 
\widetilde{p}_z^\epsilon(r) \widetilde{z}^\epsilon_\mu \widetilde{z}^\epsilon_\nu 
\end{equation}
where 
\begin{equation}\fl
\widetilde\sigma_\epsilon(r)=\frac{1}{\epsilon^2}\widetilde\sigma(r/\epsilon), \quad
\widetilde p_r^\epsilon(r)=\frac{1}{\epsilon^2}\widetilde p_\rho(r/\epsilon), \quad
\widetilde p_\phi^\epsilon(r)=\frac{1}{\epsilon^2}\widetilde p_\phi(r/\epsilon), \quad
\widetilde p_z^\epsilon(r)=\frac{1}{\epsilon^2}\widetilde p_z(r/\epsilon),
\end{equation}
and
\begin{equation}
\widetilde t^\epsilon_\mu=\sqrt\lambda \eta \widehat{\tau}^\epsilon_\mu, \quad
\widetilde r^\epsilon_\mu=\sqrt\lambda \eta \widehat{\rho}^\epsilon_\mu, \quad
\widetilde \phi^\epsilon_\mu=\sqrt\lambda \eta \widehat{\phi}^\epsilon_\mu, \quad
\widetilde z^\epsilon_\mu=\sqrt\lambda \eta \widehat{z}^\epsilon_\mu. \quad
\end{equation}
The above expression may then be used to obtain both the physical
values of the energy-momentum tensor (by taking $\epsilon=1$) and the
thin string limit (by taking $\epsilon \to 0$). 

\medskip\noindent
Before taking the thin string limit we give an expression for the mass
per unit length (taking $\epsilon=1$) in terms of $\eta^2$. 
We start by noting that
\begin{eqnarray}
\fl \mu&= 2\pi\int_0^\infty \widetilde{\sigma}(r)K(r)\d r \nonumber\\
\fl &=\pi\eta^2\int_0^\infty\left[
e^{-A}(X')^2+\frac{e^AX^2Y^2}{K^2}+2(X^2-1)^2
+\frac{1}{\alpha K^2}(Y')^2\right]K\d r. \label{mu}
\end{eqnarray}
We now complete the square on the $(X')^2$-terms and integrate by parts
to obtain
\begin{eqnarray}
\fl \mu&=\pi\eta^2\int_0^\infty\Biggl[
e^{-A}\left(X'-\frac{ne^AXY}{K}\right)^2+\frac{2nX'XY}{K}+\frac{n^2}{\alpha
K^2}(Y')^2+2(X^2-1)^2 \Biggr]K\d r \nonumber\\
\fl &=\pi\eta^2\int_0^\infty\Biggl[
e^{-A}\left(X'-\frac{ne^AXY}{K}\right)^2-\frac{nX^2Y'}{K}+\frac{n^2}{\alpha
K^2}(Y')^2+2(X^2-1)^2 \Biggr]K\d r \nonumber\\
\fl &=\pi\eta^2\int_0^\infty\Biggl\{
e^{-A}\left(X'-\frac{ne^AXY}{K}\right)^2-\frac{1}{\alpha}\left[\frac{nY'}{K}+
\frac{\alpha}{2}(X^2-1)\right]^2 \nonumber\\
\fl &-\frac{nY'}{K}+(2-\frac{\alpha}{4})(X^2-1)^2 
\Biggr\}K\d r 
\end{eqnarray}
Integrating the $Y'$-term and using the boundary conditions (\ref{C5}) for
$Y$ gives
\begin{eqnarray}
\fl \mu&=\pi n\eta^2 +\pi\eta^2\int_0^\infty\Biggl\{
e^{-A}\left(X'-\frac{ne^AXY}{K}\right)^2+\frac{1}{\alpha}\left[\frac{nY'}{K}-
\frac{\alpha}{2}(X^2-1)\right]^2 \nonumber \\
\fl &+2(1-\frac{\alpha}{8})(X^2-1)^2 
\Biggr\}K\d r. \label{mu2}
\end{eqnarray}
Note that for $\alpha$ less than the critical value of 8 the mass per
unit length is greater than $\pi n \eta^2$ whereas for $\alpha$
greater than 8 the final term in the integral is negative so that one
cannot immediately tell if $\mu$ is greater than or less than $\pi n \eta^2$.

\medskip\noindent
By adding equations (\ref{pr}) and (\ref{pphi2}), then multiplying by 
$K$, and using equation (\ref{S35}) to integrate by parts we find that
\begin{equation}
2\pi\int_0^\infty (\widetilde p_r + \widetilde p_\phi)e^A K \d r
=\frac{1}{4}\int_0^\infty(A')^2e^{-A}K\d r. \label{correct}
\end{equation}
This will be a useful equation in the next section when we relate the
angular deficit to the physical properties of the string. It is also
worth remarking at this point if we integrate $p_r+p_\phi$ over the
four dimensional region $\cal U$ bounded by the hyperplanes $t=t_0$, $t=t_0+1$
and $z=z_0$, $z=z_0+1$ we find
\begin{eqnarray}
\int_{{\cal U}}(\widetilde p_r+\widetilde p_\phi)\sqrt{(-g)} \d^4 x 
&=2\pi\int_0^\infty (\widetilde p_r+\widetilde p_\phi)e^AK\d r \nonumber\\
&=2\pi\int_0^\infty\frac{d}{dr}\left(K\frac{dA}{dr}\right)\d r \nonumber\\
&=0. \label{old} 
\end{eqnarray}
This result explains why in the flat space limit the integrated radial
and tangential pressures ($a$ and $b$) are equal and opposite, and
hence the constant $c$ in equation (\ref{cart}) in fact vanishes in
flat space. It also explains why the quantity defined by
(\ref{correct}) is very small in the weak field limit. 
This result is also consistent with the
notion that in a certain sense Cosmic Strings have no active
gravitational mass since the above result shows that the mass per unit length
defined using Tolman's expression for the mass per unit length
vanishes
\begin{eqnarray}
m&=\int_{\Re^2} (T^0_0-T^1_1-T^2_2-T^3_3)\sqrt{(-g)}\d^2x \nonumber\\
&=2\pi\int_0^\infty (\widetilde p_r+\widetilde p_\phi)e^AK\d r \quad \hbox{since $p_z=-\mu$} \nonumber\\
&=0.
\end{eqnarray}

\medskip\noindent
We now examine the thin string limit of the energy-momentum tensor. As
in the previous section we first transform to $(t,x,y,z)$ coordinates
where $x=r\cos\phi$ and $y=r\sin\phi$. Then because $\sigma_\epsilon$
takes the form
\begin{equation}
\widetilde\sigma_\epsilon(x,y)=\frac{1}{\epsilon^2}\widetilde\sigma(x,y)
\end{equation}
we can use the results of section 2 to deduce that $\sigma_\epsilon$
is associated to a delta function and
\begin{equation}
[\widetilde \sigma_\epsilon(x,y)] \approx \mu \delta^{(2)}(x,y)
\end{equation}
where $\mu$ is given by (\ref{mu}) [or (\ref{mu2})].

\medskip\noindent
As in the case of a Cosmic String in Minkowski space the
energy-momentum tensor is associated to one of the form (\ref{cart}) but
because of (\ref{old}) $c \simeq 0$ and $d \neq 0$ unless the integral of
$p_r$ (and hence $p_\phi$) vanishes. In general this does not occur
except in the case of critical coupling (see below). Therefore 
although the energy-momentum tensor is always defined as an element of
the Colombeau algebra $\Gs$ it is only associated to a distributional
energy-momentum tensor for the special case of critical coupling.

\subsection{Critical coupling}

\medskip\noindent
The field equations for a Cosmic String coupled to gravity simplify
considerably in the case of critical coupling (which with our choice
of numerical factors is given by $\alpha=8$). With this value of
$\alpha$  the final term in (\ref{mu2}) vanishes and we obtain the expression
\begin{equation}
\fl \mu=\pi n\eta^2 
+\pi\eta^2\int_0^\infty\biggl\{
e^{-A}\biggl(X'-\frac{ne^AXY}{K}\biggr)^2+\frac{1}{\alpha}\biggl[\frac{nY'}{K}-
\frac{\alpha}{2}(X^2-1)\biggr]^2
\biggr\}Kdr. \label{mu3}
\end{equation}
Since the integral is positive this gives a bound for the mass as
\begin{equation}
\mu \leqslant \pi n\eta^2.
\end{equation}
Clearly this bound will be saturated if 
\begin{equation}
A(r)\equiv 0
\end{equation}
and $X(r)$ and $Y(r)$ satisfy the curved space Bogomol'nyi equations
\begin{eqnarray}
\frac{dX}{dr}&=n\frac{XY}{K} \label{bog1}\\
\frac{dY}{dr}&=-\frac{4}{n}K(1-X^2). \label{bog2}
\end{eqnarray}
It is readily verified that if in addition $K(r)$ satisfies the
equation
\begin{equation}
K\frac{d^2K}{dr^2}=-8\pi\eta^2\left[2K^2(X^2-1)^2+n^2X^2Y^2\right]
\label{S36a}
\end{equation}
[i.e. equation (\ref{S36}) with $A=0$ and $\alpha=8$] 
then $X(r)$, $Y(r)$, $A(r)$
and $K(r)$ satisfy the field equations for a critically coupled Cosmic
String. If one imposes the boundary conditions at the origin
\begin{equation}
X(0)=0, \qquad Y(0)=1, \qquad K(0)=0, \qquad K'(0)=1,
\end{equation}
then one can look at the asymptotic behaviour of the solutions to (\ref{bog1})
and (\ref{bog2}) for large $r$ and show that in addition 
\begin{equation}
\lim_{r \to \infty}X(r)=1, \qquad \lim_{r \to \infty}Y(r)=0.
\end{equation}
So that $X(r)$, $Y(r)$, $A(r)$ and $K(r)$ given by the above equations
satisfy both the field equations (\ref{S35})--(\ref{S38}) 
and the boundary conditions (\ref{C4})--(\ref{C7}) 
for a critical Cosmic String. Conversely one can show that a
solution of the {\em critically coupled} Cosmic String field equations
is also a solution of the curved space Bogomol'nyi equations with
$A(r) \equiv 0$ and $K(r)$ a solution to (\ref{S36a}) (see Linet \cite{linet}
and Comtet and Gibbons \cite{c&g}).

\medskip\noindent
It now follows from (\ref{pr}) and (\ref{pphi2}) 
that the transverse pressures vanish
so that
\begin{eqnarray}
p_r &= 0 \\
p_\phi &=0
\end{eqnarray}
and from (\ref{mu3}) that
\begin{equation}
\mu=\pi n \eta^2.
\end{equation}
It is quite remarkable that the mass per unit length for a critically
coupled Cosmic String in a curved spacetime does not depend upon the 
details of the geometry but only on the winding number and $\eta$.
Because the transverse pressures vanish, the energy-momentum tensor of
the 1-parameter family of Cosmic Strings (which gives the thin string
limit) is associated to an ordinary  distributional energy-momentum
tensor. So that
\begin{equation}
\left[T^\epsilon_{\mu\nu}(t,x,y,z)\right] \approx \pi n \eta^2
\delta^{(2)}(x,y)\diag(1,0,0,1).
\end{equation}
Thus, as in Minkowski space, a {\em critically} coupled string can be
modelled in the thin string limit without worrying about the internal
structure (since it has a description in terms of classical
distributions) but that for a {\em non-critical} Cosmic String the
internal structure can not be completely neglected but can be
described within the Colombeau algebra (which has a finer structure
than classical distributions). 

\section{Angular deficit, holonomy and gravitational lensing}
\medskip\noindent
For the conical metric given by equation (\ref{metric1}) 
the spacetime is locally
equivalent to Minkowski space as may be seen by changing to the new
angular variable $\widetilde\phi=A\phi$. In terms of this variable the
metric becomes
\begin{equation}
ds^2=dt^2-dr^2-d\widetilde\phi^2-dz^2. \label{m2}
\end{equation}
However the spacetime is not globally flat since $0 \leqslant \widetilde\phi
\leqslant 2\pi A$. Thus (\ref{m2}) represents Minkowski space with a
wedge of angle $2\pi(1-A)$ removed and the edges identified.
In a similar way the Cosmic String spacetimes
considered in the previous section are not asymptotically flat, but
are asymptotic to conical spacetimes. A convenient mathematical way of 
measuring the angular deficit is in terms of holonomy. If one parallel
propagates an orthonormal frame once around the axis in a conical
spacetimes one finds that it has been rotated about the axis by the
deficit angle $\Delta\phi$. Thus the holonomy gives an alternative way
of measuring the angular deficit. For the Cosmic String spacetimes we
will calculate the holonomy given by parallel transport around the
curve $r=r_0$, lying in the 2-surface $t=$const. $z=$const. and then
look what happens as $r_0 \to \infty$. This will give us the angular
deficit of the limiting conical spacetime.

\medskip\noindent
Let the metric be given by
\begin{equation}
ds^2=e^A(dt^2-dr^2)-r^2e^Bd\phi^2-e^Cdz^2
\end{equation}
and let $\be_\a^\mu$ be the orthonormal tetrad given by
\begin{equation}
\be_\0^\mu=\widehat t^\mu, \qquad \be_\1^\mu=\widehat r^\mu, \qquad
\be_\2^\mu=\widehat\phi^\mu, \qquad \be_\3^\mu=\widehat z^\mu
\end{equation}
where the frame indices are in bold font to distinguish them from the
coordinate indices.
\medskip\noindent
We now consider parallel propagation around the circle
\begin{equation}
t=t_0, \qquad r=r_0, \qquad \phi \in [0, 2\pi], \qquad z=z_0.
\end{equation}
The unit tangent vector to this circle is given by $\be_\2$ and if
$\gamma^\a_{\b\c}$ denote the Ricci rotation coefficients then one may
show that
\begin{eqnarray}
\nabla_{\be_\2}\be_\1&=\gamma^\b_{\1\2}\be_\b=\gamma^\2_{\1\2}\be_\2
=r^{-1}e^{(A+B)/2}(re^{B/2})'\be_\2 \\ 
\nabla_{\be_\2}\be_\2&=\gamma^\b_{\2\2}\be_\b=\gamma^\1_{\2\2}\be_\1
=-r^{-1}e^{(A+B)/2}(re^{B/2})'\be_\1 .
\end{eqnarray}
So that a vector in the $\be_\1\wedge\be_\2$ plane remains in that plane
under parallel propagation around the circle. Let 
$\X=p(\phi)\be_\1+q(\phi)\be_2$ be a vector in the $\be_\1\wedge\be_\2$
plane then from the above
\begin{equation}
\nabla_{\be_\2}\X=(\be_\2^\mu p_{,\mu})\be_1+p\nabla_{\be_\2}\be_\1+
(\be_\2^\mu q_{,\mu})\be_\2+q\nabla_{\be_\2}\be_\2.
\end{equation}
But
\begin{equation}
\be_\2^\mu=r^{-1}e^{-B/2}\frac{\partial}{\partial \phi}
\end{equation}
and thus
\begin{equation}
\fl\hspace{18mm}\nabla_{\be_\2}\X=\frac{1}{r}e^{-B/2}\left[
\frac{dp}{d\phi}\be_\1+pe^{-A/2}(re^{B/2})'\be_\2+
\frac{dq}{d\phi}\be_\2-qe^{-A/2}(re^{B/2})'\be_\1 \right]
\end{equation}
Hence $\nabla_{\be_\2}\X=0$ implies
\begin{eqnarray} 
\frac{dp}{d\phi}&=\kappa q \\
\frac{dq}{d\phi}&=-\kappa p 
\end{eqnarray}
where
\begin{equation}
\kappa=e^{-A/2}\frac{d}{dr}\left(re^{B/2}\right)|_{r=r_0}.
\end{equation}
Solving this equation gives
\begin{eqnarray}
p(2\pi)&=\cos(2\pi\kappa)p(0)-\sin(2\pi\kappa)q(0) \\
q(2\pi)&=\sin(2\pi\kappa)p(0)+\cos(2\pi\kappa)q(0)
\end{eqnarray}
so that the vector $\X$ rotates by $-2\pi\kappa$ relative to the frame
$\be_\1$, $\be_\2$. On the other hand this frame rotates by $2\pi$ as it
goes round the loop, and  $\X$ is rotated by $2\pi(1-\kappa)$ as
a result of parallel propagation around the circle $r=r_0$. The
holonomy at infinity gives the angular deficit and hence
\begin{equation}
\Delta\phi=\lim_{r \to \infty}2\pi
\left[1-e^{-A/2}\frac{d}{dr}\left(re^{B/2}\right)\right]
\end{equation}
or in terms of $A$ and $K$
\begin{equation}
\Delta\phi=\lim_{r \to \infty}2\pi
\left[1-e^{-A/2}\frac{d}{dr}\left(Ke^{-A/2}\right)\right].
\end{equation}
\medskip\noindent
Because the $t=$const. $z=$const. 2-plane has zero extrinsic curvature
the holonomy is also the same as that produced by the connection of
the induced 2-metric. The holonomy in two dimensions may also be
computed using the Gauss-Bonnet theorem (see eg Vickers \cite{JAV})
and is given by
\begin{equation}
Hol(D_r)=\int_{D_r} \widetilde \Omega^1_{\ 2}=\int_{D_r}\Omega^1_{\ 2}
\end{equation}
where $D_r$ is the disk of radius $r$, $\widetilde\Omega^1_{\ 2}$ is the
curvature 2-form in the 2-plane and $\Omega^1_{\ 2}$ is the curvature
2-form of the 2-plane in the spacetime (which are the same since the
extrinsic curvature vanishes). Taking the limit as $r \to \infty$
gives
\begin{equation}
\Delta\phi=\int_{\Re^2} \Omega^1_{\ 2}.
\end{equation}
Using the result of Proposition 3 in the Appendix we obtain
\begin{equation}
\Delta\phi=8\pi\mu+2\pi(P_r+P_\phi) \label{result}
\end{equation}
where
\begin{equation}
P_r=2\pi\int_0^\infty\widetilde p_r K \d r ,
\end{equation}
and
\begin{equation}
P_\phi=2\pi\int_0^\infty\widetilde p_\phi K \d r .
\end{equation}
Note that equation (\ref{correct}) shows that this is consistent with
Garfinkle's result (equation (82) of \cite{garfinkle}). Using
equations (\ref{sigma2}), (\ref{pr}) and (\ref{pphi2}) 
we may obtain an explicit
expression for $\Delta\phi$ in terms of the matter fields similar to
equation (\ref{mu}) for $\mu$. 
\begin{equation}
\fl\hspace{5mm} \Delta\phi=8\pi^2\eta^2\int_0^\infty\left[e^A\left(\frac{dX}{dr}\right)^2
+n^2e^A\frac{X^2Y^2}{K^2}+(X^2-1)+\frac{3n^2}{2\alpha
K^2}\left(\frac{dY}{dr}\right)\right]K\d r.
\end{equation}
However the significant thing for us is
that (\ref{result}) gives an expression for the angular deficit in
terms of the integrated matter fields. 

\medskip\noindent
If one now considers the 1-parameter family of metrics given by
Proposition 2, which give the thin string limit, one finds that {\em
all} the terms in (\ref{result}) are independent of $\epsilon$ so that
(\ref{result}) is true for {\em both} the physical spacetime and the
thin string limit spacetime. One may therefore give an expression for
the angular deficit of the physical spacetime in terms of the
distributional energy-momentum tensor of the thin string limit. Note
that this result is not incompatible with  Futamase and Garfinkle's claim
that the thin string limit is not well defined using conventional
distribution theory since the $P_r$ and $P_\phi$ terms are not well
defined unless one uses a theory such as Colombeau algebras which
extends the concept of distribution.

\medskip\noindent
The calculation above shows how the deficit angle $\Delta\phi$ can be
obtained in terms of the matter distribution. However the deficit
angle can not be directly measured so we now consider the deflection
of a light ray due to the gravitational field of a Cosmic String. One
may calculate this by looking at the equations for the null
geodesics. For the metric given by (\ref{metric}) these may be calculated using 
the Euler-Lagrange equations for the Lagrangian
\begin{equation}
{\L}
=\half(e^a\dot\tau^2-e^a\dot\rho^2-\rho^2e^b\dot\phi^2-e^a\dot
z^2)
\end{equation}
We will consider geodesics moving in the plane orthogonal to the
string so that $z=\mathrm{const}.$ and $\dot z=0$. We now use the absence of
explicit $\tau$ and $\phi$ dependence to show that we have two
conserved quantities:
\begin{eqnarray}
\frac{\partial{\L}}{\partial{\dot \tau}}&=e^a\dot\tau=E=\mathrm{const}. \\
\frac{\partial{\L}}{\partial{\dot \phi}}&=-\rho^2e^b\dot\phi=J=\mathrm{const}.
\end{eqnarray}

\medskip\noindent
One may then calculate the scattering angle for a pair of geodesics
which pass either side of the string. One finds that in these
coordinates the angular deflection $\phi_d$ is given by
\begin{equation}
\phi_d=2\pi-\frac{4E}{J}\int_{\rho_c}^\infty
\frac{\d\rho}{g(\rho)[g(\rho)^2-1]^{1/2}} \label{phid}
\end{equation}
where $g(\rho)=\frac{E}{J}e^{\frac{1}{2}(b-a)}$ and $\rho_c$ is the closest
approach of the light ray to the $\rho=0$ axis (which satisfies the
condition $g(\rho_c)=1$).

\medskip\noindent
In general we see that the scattering depends upon $I=E/J$ which is
essentially just the impact parameter for the geodesic. However for
the special case of a conical metric
\begin{equation}
ds^2=d\tau^2-d\rho^2-A^2\rho^2d\phi^2-dz^2
\end{equation}
One can show that the scattering is independent of $I$. More precisely
one finds that
\begin{equation}
\phi_d=2\pi-\frac{4}{A}\int_1^\infty\frac{r\d r}{r^2(r^2-1)^{1/2}}
=\frac{2\pi(1-A)}{A}
\end{equation}

\medskip\noindent
If one now transforms to the $\widetilde\phi$ coordinates one finds
\begin{equation}
\widetilde\phi_d=A\phi_d=2\pi(1-A)=\Delta\phi
\end{equation}
So that in the coordinates in which the cone is written as Minkowski
space minus a wedge, the scattering angle is equal to the angular
deficit.

\medskip\noindent
If one considers a possible observation of a multiple image of a star
due to gravitational lensing by a Cosmic String then the closest
approach of the geodesic would be approximately given by
$\rho_c \sim R\Delta\phi$, where $R$ is the distance to the star. Thus
the ratio of the closest approach to the thickness $\rho_c/\rhoo$
would be very large. This implies that the impact parameter would be large
compared to the thickness of the string. Thus when looking at the
effect of gravitational lensing in the thin string limit the
appropriate limit to consider is to take $\epsilon \to 0$ while
keeping the impact parameter $I$, constant. We call this the {\it
lensing limit}. In calculating the lensing limit the deflection takes
place in a region which gets closer and closer to a conical spacetime
as $\epsilon \to 0$. One can use this to show that in the {\it lensing
limit} the scattering is independent of the impact parameter and is
equal to that of a cone with deficit angle $\Delta\phi$ given by
equation (\ref{result}). 

\medskip\noindent
However as well as the lensing limit one can also examine the limit in
which the closest approach $\rho_c$ scales with the thickness of the
string so that $\rho_c/\rhoo$ is independent of $\epsilon$. We call
this limit the {\it grazing limit}. In this case the scattering angle
turns out to be independent of $\epsilon$ and is given by (\ref{phid}) for all
values of $\epsilon$. The deflection in this case does of course
depend upon the impact parameter, but a suitable choice to consider is
$I=1$ since this gives the angle between to geodesics which are
initially parallel and with a separation approximately equal to the
effective diameter of the string. The results are plotted in Figure 1
which show $\phi_d$ and $\Delta\phi$ as a function of $\alpha$
and $\eta$ respectively.

    \begin{figure}[H] \centering
    \mbox{\subfigure[$\phi_d$ and $\Delta\phi$ $\mathrm{[deg.]}$ as a function of $\alpha$ for $\eta=0.1$.]{\includegraphics[scale=0.45]{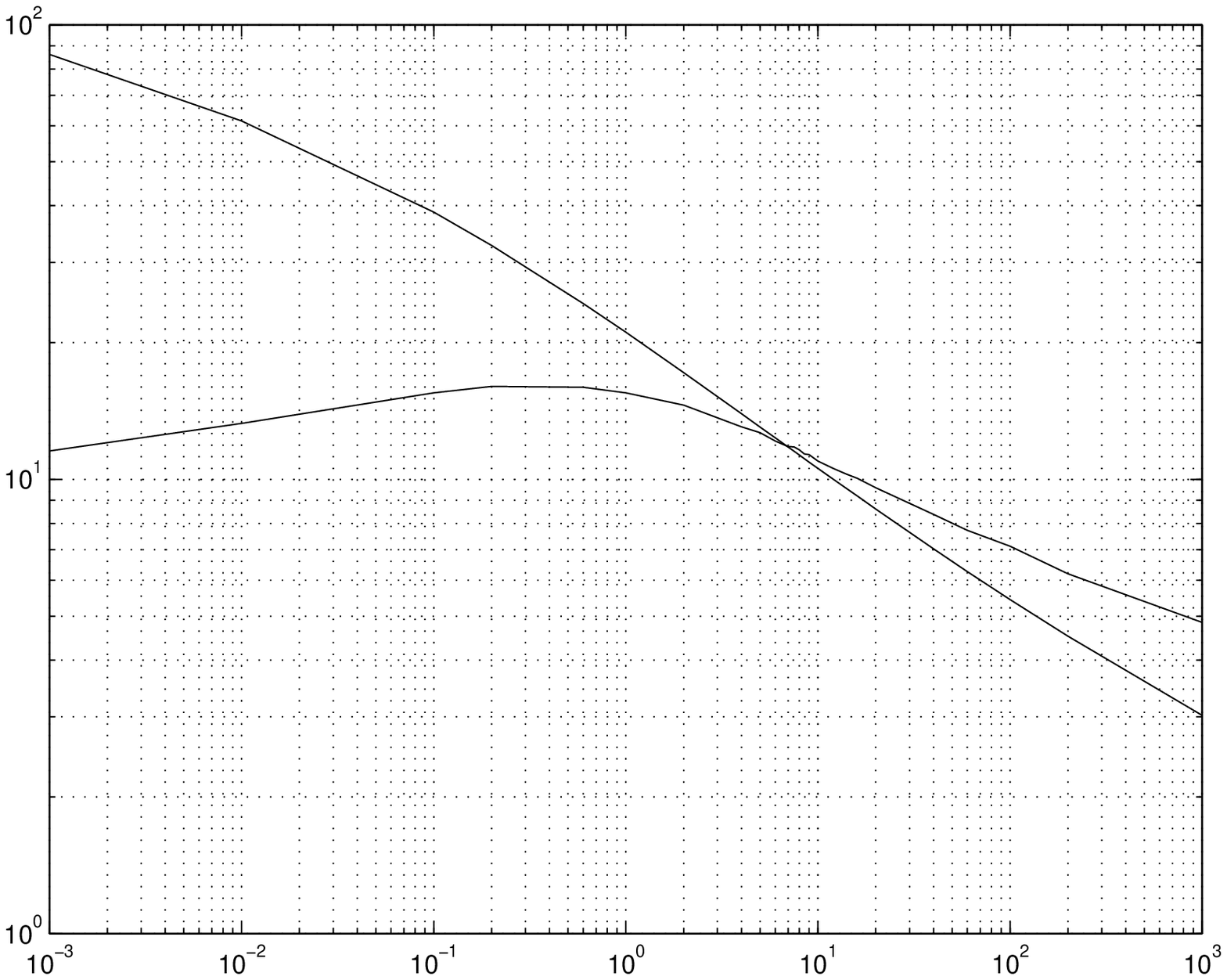}}\quad
          \subfigure[$\phi_d$ and $\Delta\phi$ $\mathrm{[deg.]}$  as a function of $\eta$ for $\alpha=1$.]{\includegraphics[scale=0.45]{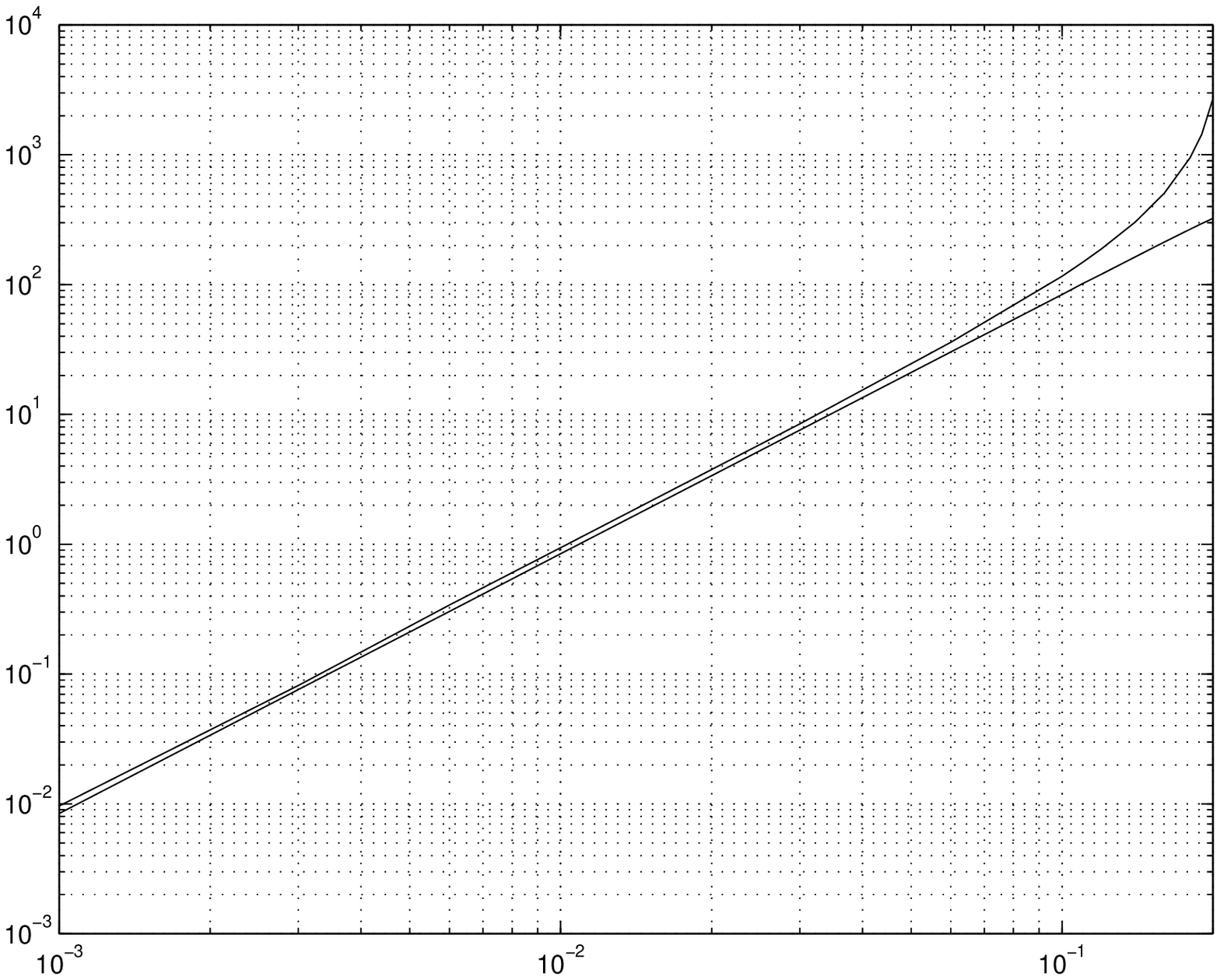}}}
    \put(-103,-3){\fontsize{8}{8pt}\selectfont\makebox(0,0)[t]{$\eta$}}
    \put(-332,-3){\fontsize{8}{8pt}\selectfont\makebox(0,0)[t]{$\alpha$}}
    \put(-370,115){\fontsize{8}{8pt}\selectfont\makebox(0,0)[t]{$\phi_d$}}
    \put(-370,150){\fontsize{8}{8pt}\selectfont\makebox(0,0)[t]{$\Delta\phi$}}
    \put(-40,110){\fontsize{8}{8pt}\selectfont\makebox(0,0)[t]{$\phi_d$}}
    \put(-40,130){\fontsize{8}{8pt}\selectfont\makebox(0,0)[t]{$\Delta\phi$}}
    \caption{Scattering angle compared to deficit angle \index{deficit angle} as a function of $\alpha$ and $\eta$ respectively. Notice that $\phi_d=\Delta\phi$ for $\alpha=8$.}
    \end{figure}

\section{The Conical approximation}
\medskip\noindent
We now wish to consider the asymptotic behaviour as $\epsilon \to 0$ 
of a 1-parameter family of solutions representing the thin string
limit. As a result of proposition 2 we know that both the string and
metric variables scale according to
\begin{equation}
 f_\epsilon(\rho)=f(\rho/\epsilon) .
\end{equation}
Thus in the `far zone' given by $\rho > \epsilon$ 
the asymptotic behaviour of $f_\epsilon(\rho)$
is given by the asymptotic behaviour of $f(r)$ for large $r$. This
gives the leading order behaviour of the metric variables as 
\begin{eqnarray} 
A(r) &\sim A_0 \\
B(r) &\sim B_0
\end{eqnarray}
where $A_0$ and $B_0$ are constants such that 
\begin{equation}
ds^2=e^{A_0}(dt^2-dr^2)-r^2e^{B_0}dr^2-e^{A_0}dz^2
\end{equation}
is the metric of a conical spacetime with deficit angle $\Delta\phi$
given by (\ref{result}). 
Calculating the higher order asymptotic behaviour shows
that the metric variables have asymptotic expansions of the form
\begin{eqnarray}
A(r) &=A_0+\frac{A_1}{r}+O(\frac{1}{r^2}) \\
B(r) &=B_0+\frac{B_1}{r}+O(\frac{1}{r^2}) .
\end{eqnarray}
In contrast to the asymptotic expansion of the metric variables in 
inverse powers of $r$ one finds that the matter variables have
exponentially fast decay. If one defines $Z(r)=X(r)-1$ and
$W(r)=Y(r)/r$ then to leading order $X(r)$ and $Y(r)$ satisfy the
equations
\begin{eqnarray}
r\frac{d}{dr} \left(r\frac{dZ}{dr}\right)-8e^{A_0}r^2Z&=0 \\
r\frac{d}{dr} \left(r\frac{dW}{dr}\right)-(1+\alpha e^{A_0}r^2)W&=0 \ .
\end{eqnarray}
These are hyperbolic Bessel equations of order 0 and 1 respectively
and give the asymptotic behaviour as
\begin{eqnarray}
Z(r) &\sim K_0(c_1r) \sim r^{-1/2}e^{-c_1r} \\
W(r) &\sim K_1(c_2r) \sim r^{-1/2}e^{-c_2r} 
\end{eqnarray}
where $c_1=2\sqrt 2 e^{A_0/2}$ and $c_2=\sqrt\alpha e^{A_0/2}$.

\medskip\noindent
Calculating higher order terms leads to asymptotic expansions for
$X(r)$ and $Y(r)$ of the form
\begin{eqnarray}
X(r)&=1+r^{-1/2}e^{-c_1r}\left[X_0+\frac{X_1}{r}+O(\frac{1}{r^2})\right] \\
Y(r)&=r^{1/2}e^{-c_2r}\left[Y_0+\frac{Y_1}{r}+O(\frac{1}{r^2})\right]
\ .
\end{eqnarray}
Because of the rapid exponential decay of the matter terms compared
with the polynomial in $1/r$ decay of the metric variables, one finds
that even in the `near zone' ($\rho < \epsilon$) provided one remains
outside the exponentially small inner `core zone' then 
the asymptotic behaviour of the metric is the same as that of a
cylindrically symmetric {\it vacuum} spacetime which is asymptotic to 
a conical spacetime.  We therefore introduce the `conical
approximation' in which we model the matter as a conical singularity
with angular deficit given by (\ref{result}) and consider the
gravitational field as a perturbation of this metric. Because the
gravitational field satisfies a linear field equation in this
situation one can solve for the perturbation \emph{exactly} in terms
of a cylindrically symmetric solution of Laplace's equation and then
write the gravitational field as a superposition of this with the
conical solution.

\medskip\noindent
The asymptotic behaviour for the above variables for large $r$
translates into the following thin string
limit outside the core.
\begin{eqnarray}
a_\epsilon(\rho) &= a_0 +\frac{\epsilon a_1}{\rho}+O(\epsilon^2) \\   
b_\epsilon(\rho) &= b_0 +\frac{\epsilon b_1}{\rho}+O(\epsilon^2) \\   
S_\epsilon(\rho) &= \eta
+\epsilon^{1/2}\rho^{-1/2}e^{-c_1\rho/\epsilon}
\left[S_0+\frac{\epsilon S_1}{\rho}+O(\epsilon^2)\right] \\   
P_\epsilon(\rho) &= 
\epsilon^{-1/2}\rho^{1/2}e^{-c_2\rho/\epsilon}
\left[P_0+\frac{\epsilon P_1}{\rho}+O(\epsilon^2)\right] \ .    
\end{eqnarray}

\section{The thin string limit for dynamic Cosmic Strings}

In this section we consider the thin string limit for a time
dependent Cosmic String. We start with the Lagrangian (\ref{lag}) and with
$\Phi$ and $A_\mu$ given by (\ref{S6a}) and (\ref{S6b}) 
but now take $S$ and $P$ to be
functions of both $\tau$ and $\rho$. We first consider the case of a
dynamic Cosmic String on a fixed Minkowski background. The equations
of motion are then given by
\begin{eqnarray}
\rho\frac{\partial}{\partial\rho}
\biggl(\rho\frac{\partial S}{\partial\rho}\biggr)
-\rho^2\frac{\partial^2S}{\partial \tau^2} 
= S[4\lambda\rho^2(S^2-\eta^2)+P^2] \label{T1}\\
\rho\frac{\partial}{\partial\rho}
\biggl(\rho^{-1}\frac{\partial P}{\partial\rho}\biggr)
-\frac{\partial^2P}{\partial \tau^2}  
= e^2S^2P. \label{T2}
\end{eqnarray}
We now introduce the Colombeau thin string parameter $\epsilon$  and
use this to scale the coupling constants $e$, $\lambda$ and $\eta$
according to equations (\ref{ib})--(\ref{iiib}). 
We then consider 1-parameter families of
solutions $S_\epsilon$ and $P_\epsilon$  to the equations
\begin{eqnarray}
\rho\frac{\partial}{\partial\rho}
\biggl(\rho\frac{\partial S_\epsilon}{\partial\rho}\biggr)
-\rho^2\frac{\partial^2S_\epsilon}{\partial \tau^2} 
= S_\epsilon[4\lambda_\epsilon\rho^2(S^2_\epsilon-\eta^2)+P^2_\epsilon] \label{T3}\\
\rho\frac{\partial}{\partial\rho}
\biggl(\rho^{-1}\frac{\partial P_\epsilon}{\partial\rho}\biggr)
-\frac{\partial^2P_\epsilon}{\partial \tau^2}  
= e_\epsilon^2 S^2_\epsilon P_\epsilon. \label{T4}
\end{eqnarray}
The solutions to (\ref{T3}) and (\ref{T4}) are given by
\begin{eqnarray}
S_\epsilon(\tau, \rho)&=\eta X(\sqrt\lambda\eta\rho/\epsilon, 
\sqrt\lambda\eta\tau/\epsilon) \label{T3a}\\
P_\epsilon(\tau, \rho)&= Y(\sqrt\lambda\eta\rho/\epsilon, 
\sqrt\lambda\eta\tau/\epsilon) \label{T4a}
\end{eqnarray}
where $X(t,r)$ and $Y(t,r)$ are solutions to 
\begin{eqnarray}
r\frac{\partial}{\partial r}\biggl(r\frac{\partial X}{\partial r}\biggr) 
-r^2\frac{\partial^2X}{\partial t^2}
=X[4 r^2(X^2-1)+Y^2] \label{T5}\\
r\frac{\partial}{\partial r}\biggl(r^{-1}\frac{\partial Y}{\partial r}\biggr) 
-\frac{\partial^2Y}{\partial t^2}
= \alpha X^2Y. \label{T6}
\end{eqnarray}
We see from the form of $P_\epsilon$ and $S_\epsilon$ given by
(\ref{T3a}) and (\ref{T4a}) that taking the limit as
$\epsilon \to 0$ corresponds to simultaneously taking a limit in which
both $\rho \to \infty$ and $\tau \to \infty$. Since a dynamic Cosmic
String eventually settles down to a static configuration taking the 
thin string limit is also a limit in which
the string also becomes static.

\medskip\noindent
In order to investigate the asymptotic behaviour for small epsilon
(for $\rho \neq 0$) we consider solutions of (\ref{T5}) and (\ref{T6})
for large values of $r$. These may be written  
\begin{eqnarray}
X&=1+X_1 \\
Y&=Y_1
\end{eqnarray}
where $X_1$ and $Y_1$ are small for large $r$. Substituting into
(\ref{T5}) and (\ref{T6}) and ignoring terms which are quadratic or
higher in $X_1$ and $Y_1$ we find they satisfy
\begin{eqnarray}
r\frac{\partial}{\partial r}\biggl(r\frac{\partial X_1}{\partial r}\biggr) 
-8 r^2X_1-r^2\frac{\partial^2X_1}{\partial t^2}&=0 \label{T7} \\
r\frac{\partial}{\partial r}\biggl(r^{-1}\frac{\partial Y_1}
{\partial r}\biggr) 
-\alpha Y_1-\frac{\partial^2Y_1}{\partial t^2}
&=0. \label{T8}
\end{eqnarray}
We may solve (\ref{T7}) by considering a solution of the form
\begin{equation}
X_\omega(t,r)=Z(r)e^{i\omega t}
\end{equation}
and substituting into (\ref{T7}) to obtain
\begin{equation}
r\frac{\partial}{\partial r}\biggl(r\frac{\partial Z}{\partial r}\biggr) 
-(8\lambda-\omega^2)r^2Z=0 \ .\label{T9} 
\end{equation}
The solution to this which is finite as $r \to \infty$ is given by
$Z(r)=K_0(c_1 r)$ where $c_1^2=8-\omega^2$. Hence
\begin{equation}
X_\omega(t,r)=K_0(\sqrt{8-\omega^2}r)e^{i\omega t}
\end{equation}
and the general solution to (\ref{T7}) may be written
\begin{equation}
X(t,r)=\int A_\omega K_0(\sqrt{8-\omega^2}r)e^{i\omega t} d\omega \ .
\end{equation}
In terms of the physical variables this gives the asymptotic behaviour
in the far zone ($\rho > \epsilon$) as $\epsilon \to 0$ as
\begin{equation}
S_\epsilon(\rho, \tau) \sim \eta + \int A_\omega 
K_0(\sqrt{(8\lambda-\omega^2)\lambda\eta^2}\rho/\epsilon)
e^{i\sqrt\lambda\eta\omega\tau/\epsilon} d\omega \ .\label{freq1}
\end{equation}
In a similar way one can show that
\begin{equation}
P_\epsilon(\rho, \tau) \sim \frac{\rho}{\epsilon}\int B_\omega 
K_1(\sqrt{(e^2\eta^2-\omega^2)\lambda\eta^2}\rho/\epsilon)
e^{i\sqrt\lambda\eta\omega\tau/\epsilon} d\omega \ .\label{freq2}
\end{equation}
The important thing to note is that $S_\epsilon$ and $P_\epsilon$
still have an exponential decay as $\epsilon \to 0$ 
due to the exponential decay of $K_0$ and $K_1$ for large $r$ so these
expansions remain valid in the near zone provided one stays outside
the exponentially small inner core.

\medskip\noindent
A similar analysis of a Cosmic String coupled to the gravitational
field shows that $S_\epsilon$ and $P_\epsilon$ again have the
$e^{-c\rho/\epsilon}$ type exponential fall off. By contrast the
metric coefficients have asymptotic expansions which are polynomial in
$\epsilon$. In the dynamic case we may therefore also introduce the
conical approximation in which we model the matter as a static conical
singularity with angular deficit given by (\ref{result}) and consider
the gravitational field as a perturbation of this solution. Because of
the linearity of the vacuum Einstein equations for a cylindrically
symmetric metric written in Rosen form, one can explicitly write down
the solution as a superposition of a conical solution with a solution
of the cylindrical wave equation. Such a situation has been analysed
by Marder \cite{marder}. 

Another important feature of (\ref{freq1}) and (\ref{freq2}) is that
they may be used to determine the frequency of oscillation of the
matter fields. Since $K_0$ and $K_1$ decay exponentially fast, the
dominant contribution to the frequency is when the argument is close
to zero. This gives the frequencies of the fields $S$ and $P$ as 
\begin{eqnarray}
f_S&=\sqrt{2\lambda} \eta /\pi \\
f_P&= e \eta /2\pi \ .
\end{eqnarray}
A plot of the corresponding values for the rescaled fields $X$ and
$Y$, compared
to the numerically computed values (using the results of \cite{SSV2})
is shown in Figure 2.

    \begin{figure}[H] \centering
    \mbox{\subfigure{\includegraphics[scale=0.45]{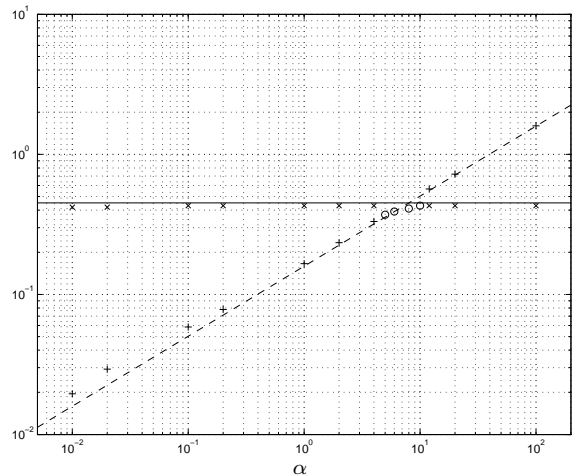}}}
    \put(-104,-3){\fontsize{8}{8pt}\selectfont\makebox(0,0)[t]{$\alpha$}}
    \caption{The frequecies $f_X$ and $f_Y$ calculated using the
    asymptotics compared to the numerically computed values.}
    \end{figure}

\section{Conclusion}

In this paper we have examined the thin string limit of a Cosmic String
coupled to the gravitational field. By using a description in terms of
the simplified Colombeau algebra we have been able to show that the
energy-momentum tensor has a well defined limit. Furthermore we have
shown how the angular deficit of the limiting conical spacetime may be
expressed in terms of the limiting energy-momentum tensor through
equation (\ref{result}). However it is important to note that in
general the Colombeau distributional energy-momentum tensor that one
obtains is in general not associated to a conventional
distribution. It is only in the case of critical coupling that the
energy-momentum tensor can be written in terms of Dirac delta
functions. Thus the results we obtain are not in conflict with those
of Garfinkle for example \cite{garfinkle} who showed that there were
problems describing the thin string limit. It is only by using the
features of the Colombeau algebra which allow one to describe the
internal structure of the line source that it is possible to describe
the thin string limit. For example a limiting distributional 
tangential stress cannot be defined using conventional distributions
but can be described in terms of the Colombeau algebra.

\medskip\noindent
In the last two sections we examined the asymptotic behaviour of both
the metric and matter variables in the thin string limit. We showed
that in the far zone the metric variables have polynomial decay while
the matter variables decay exponentially. We were therefore able to
introduce in both the static and dynamic case the notion of the
conical approximation. In this approximation one replaces the matter
terms by a line source with distributional energy-momentum tensor and
a conical singularity which has angular deficit given in terms of the
matter through (\ref{result}). One is then able to model the
gravitational field as a solution of the vacuum Einstein equations with    
the required angular deficit on the axis. Because of the linearity of
the field equations one can solve this explicitly by looking at
cylindrically symmetric solutions of the wave equation which are
regular on the axis and looking at a superposition of this with 
a static solution representing a conical spacetime. Because of the
exponential fall off of the matter fields this approximation is valid
even in the near zone as long as one remains outside the exponentially
small inner core of the string. We have shown how this approximation
may be used to calculate the frequency of oscillation of the matter
fields and this shows very good agreement with the numerically
calculated frequencies of \cite{SSV2}. In a future paper this idea
will be developed and the conical
approximation will be used to examine further properties of a dynamic
Cosmic String and the results compared to numerical calculations.

\section{Appendix: A formula for the deficit angle in terms of
integral quantities}
In this appendix we obtain an expression for the angular deficit in
terms of integrals of the components of the energy-momentum tensor. 
The calculation just uses the Einstein tensor and is entirely in terms of
the metric coefficients. No use is made of the field equations for the
matter variables.
\medskip\noindent
In this section we take the metric to be given by
\begin{equation}
ds^2=e^{2a}dt^2-e^{2b}dr^2-e^{2c}d\phi^2-e^{2a}dz^2
\end{equation}
(Note the factors of 2 in the exponents compared to earlier sections.)
and let $\be_\a^\mu$ be the orthonormal tetrad given by
\begin{equation}
\be_\0^\mu=\widehat t^\mu, \qquad \be_\1^\mu=\widehat r^\mu, \qquad
\be_\2^\mu=\widehat\phi^\mu, \qquad \be_\3^\mu=\widehat z^\mu
\end{equation}
where the frame indices are in bold font to distinguish them from the
coordinate indices.
Then the calculations of section 5 show that the angular deficit is
given by
\begin{equation}
\Delta\phi=\lim_{r \to \infty}2\pi(1-c'e^{c-b}).
\end{equation}
Using the Gauss-Bonnet theorem one can also write this as an integral
\begin{equation}
\Delta\phi=-2\pi\int_0^\infty [c''+(c')^2-c'b']e^{c-b}\d r.
\end{equation}
On the other hand using $\sigma=T_{\0\0}=1/(8\pi)G_{\0\0}$ (where the
use of bold font indicates the use of frame indices) we get
\begin{equation}
\sigma=-1/(8\pi)[a''+c''+(a')^2+(c')^2-b'c'-a'b'+a'c']e^{-2b}. 
\end{equation}
So that the mass per unit length is given by
\begin{eqnarray}
\mu&=\int_0^{2\pi}\int_0^\infty \sigma \sqrt{{}^{(2)}g} \d r\d\phi \\
   &= 2\pi \int_0^\infty \sigma e^{b+c} \d r \\
   &= -\frac{1}{4}\int_0^\infty
   [a''+c''+(a')^2+(c')^2-b'c'-a'b'+a'c']e^{c-b}\d r. 
\end{eqnarray}
But 
\begin{eqnarray}
\int_0^\infty a''e^{c-b}\d r
&=[a'e^{c-b}]_0^\infty-\int_0^\infty(a'c'-a'b')e^{c-b}\d r \\
&=-\int_0^\infty(a'c'-a'b')e^{c-b}\d r. 
\end{eqnarray}
Hence
\begin{equation}
\mu=-\frac{1}{4}\int_0^\infty [c''+(c')^2-b'c'+(a')^2]e^{c-b}\d r
\end{equation}
and thus
\begin{eqnarray}
8\pi\mu&=-2\pi\int_0^\infty [c''+(c')^2-b'c']e^{c-b}\d r
                -2\pi\int_0^\infty(a')^2 e^{c-b}\d r \\
      & = \Delta\phi-2\pi\int_0^\infty(a')^2 e^{c-b}\d r. 
\end{eqnarray}
\medskip\noindent
We now show how to write this entirely in terms of integrated
components of the energy-momentum tensor. Using
$p_r=T_{\1\1}=1/(8\pi)G_{\1\1}$ we have
\begin{equation}
p_r= 1/(8\pi)[(a')^2+2a'c']e^{-2b} 
\end{equation}
Let $P_r$ be the integrated version (with respect to the volume form
determined by the determinant of the 2-metric on t=const. z=const.)
Then we have 
\begin{eqnarray}
P_r&=\int_0^{2\pi}\int_0^\infty p_r e^{c+b}\d r\d\phi \\
   &= \frac{1}{4}\int_0^\infty [(a')^2+2a'c']e^{c-b}\d r. 
\end{eqnarray}
Similarly using $p_\phi=T_{\2\2}=1/(8\pi)G_{\2\2}$ we have 
\begin{equation}
p_\phi=1/(8\pi)[2a''+3(a')^2-2a'b']e^{-2b} 
\end{equation}
so that the integrated version $P_\phi$ is given by
\begin{equation}
P_\phi=\frac{1}{4}\int_0^\infty [2a''+3(a')^2-2a'b']e^{c-b}\d r.
\end{equation}
But (as before)
\begin{eqnarray}
\int_0^\infty a''e^{c-b}\d r
&=[a'e^{c-b}]_0^\infty-\int_0^\infty(a'c'-a'b')e^{c-b}\d r \\
&=-\int_0^\infty(a'c'-a'b')e^{c-b}\d r 
\end{eqnarray}
so that
\begin{equation}
P_\phi=\frac{1}{4}\int_0^\infty [3(a')^2-2a'c']e^{c-b}\d r.
\end{equation}
Thus
\begin{equation}
P_r+P_\phi= \int_0^\infty (a')^2 e^{c-b}\d r.
\end{equation}
So that
\begin{equation}
\Delta\phi=8\pi\mu+2\pi(P_r+P_\phi).
\end{equation}
\medskip\noindent
We have thus established the following proposition.
\begin{proposition}
The angular deficit $\Delta\phi$ for the metric given by
\begin{equation}
ds^2=e^{2a}dt^2-e^{2b}dr^2-e^{2c}d\phi^2-e^{2a}dz^2
\end{equation}
where $a$, $b$ and $c$ are only functions of $r$ is given by
\begin{equation}
\Delta\phi=8\pi\mu+2\pi(P_r+P_\phi)
\end{equation}
where $\mu$, $P_r$ and $P_\phi$ are the integrated values of 
$1/(8\pi)G_{\0\0}$, $1/(8\pi)G_{\1\1}$ and
$1/(8\pi)G_{\2\2}$ respectively, and the integration is with respect
to the volume form induced on the 2-surface $t=\mathrm{const}$. $z=\mathrm{const}$.
\end{proposition}


\Bibliography{15}


\bibitem{vilenkin} Vilenkin A., Gravitational field of vacuum domain walls and strings, \emph{Phys. Rev. D}, \textbf{23}, 852 (1981)

\bibitem{hiscock} Hiscock W.A., Exact gravitational field of a string,
\emph{Phys. Rev. D} \textbf{31}, 3288 (1985) 

\bibitem{g&t} Geroch R.P. and Traschen J., Strings and other distributional sources in General Relativity, \emph{Phys. Rev. D}, \textbf{38}, 1017 (1987)

\bibitem{garfinkle} Garfinkle D., General relativistic strings, \emph{Phys. Rev. D}, \textbf{32}, 1323 (1985)

\bibitem{israel} Israel W., Line sources in general relativity,
\emph{Phys. Rev. D}, \textbf{15}, 935 (1977)

\bibitem{linet} Linet B., The Static Metrics with Cylindrical Symmetry Describing a Model of Cosmic Strings, \emph{Gen. Rel. and Grav.}, \textbf{17}, 1109 (1985)

\bibitem{futamase}  Futamase T. and Garfinkle D., What is the relation between $\Delta\phi$ and $\mu$ for a cosmic string?, \emph{Phys. Rev. D}, \textbf{37}, 2086 (1988)

\bibitem{wilson} Wilson J.P., \emph{Regularity of Axisymmetric Space-times in General Relativity}, University of Southampton, 1998

\bibitem{cvw} Clarke C.J.S., Vickers J.A. and Wilson J.P., Generalized functions and distributional curvature of cosmic strings, \emph{Class. Quantum Grav.}, \textbf{13}, 2485 (1996)

\bibitem{col1} Colombeau J.F., \emph{Multiplication of Distributions}, Springer-Verlag, 1992

\bibitem{biagioni} Biagioni H.A., \emph{A nonlinear theory of
generalised functions} Lecture Notes in mathematics 1421, Springer, 1990

\bibitem{advances} Grosser M, Kunzinger M, Steinbauer R and Vickers
J.A., A global theory of algebras of generalized functions
\emph{Advances in Mathematics} to appear

\bibitem{mo} Oberguggenberger M., \emph{Multiplication of distributions
and applications to partial differential equations}, Pitman Research
Notes in Mathematics 259, Longman, 1992

\bibitem{shellard} Shellard E.P.S. and Vilenkin A., \emph{Cosmic Strings and other Topological Defects}, Cambridge, 1994

\bibitem{SSV1} Sj\"odin K.R.S., Sperhake U. and Vickers J.A., Dynamic
cosmic strings I, \emph{Phys. Rev. D}, \textbf{63}, 024011 (2001)

\bibitem{clarke&wilson} Wilson J.P. and Clarke C.J.S., Elementary
flatness on a symmetry axis, \emph{Class. Quantum Grav.}, \textbf{13}, 
2007 (1996)

\bibitem{c&g} Comtet A. and Gibbons G.W., Bogomol'nyi bounds for
cosmic strings \emph{Nucl. Phys. B} \textbf{299}, 719 (1988)

\bibitem{JAV} Vickers J.A.G., Generalised cosmic strings, \emph{Class. Quantum Grav.}, \textbf{4}, 1 (1987)

\bibitem{marder} Marder L., On the Existence of Cylindrical Gravitational Waves, Ph.D. thesis, 1957

\bibitem{SSV2} Sperhake U., Sj\"odin K.R.S. and Vickers J.A., Dynamic
cosmic strings II, \emph{Phys. Rev. D}, \textbf{63}, 024012 (2001)

\endbib
\end{document}